# Spread, volatility, and volume relationship in financial markets and market maker's profit optimization

Jack Sarkissian

Managing Member, Algostox Trading LLC

email: jack@algostox.com

Abstract

We study the relationship between price spread, volatility and trading volume. We find that spread forms as a result of interplay between order liquidity and order impact. When trading volume is small adding more liquidity helps improve price accuracy and reduce spread, but after some point additional liquidity begins to deteriorate price. The model allows to connect the bid-ask spread and high-low bars to measurable microstructural parameters and express their dependence on trading volume, volatility and time horizon. Using the established relations, we address the operating spread optimization problem to maximize the market-maker's profit.

## 1. Introduction

When discussing security prices it is customary to describe them with single numbers. For example, someone might say price of Citigroup Inc. (ticker "C") on April 18, 2016 was \$45.11. While good enough for many uses, it is not entirely accurate. Single numbers can describe price only as referring to a particular transaction, in which $N$ units of security are transferred from one party to another at a price \$$X$ each. Price could be different a moment before or after the transaction, or if the transaction had different size, or if it were executed on a different exchange. To be entirely accurate, one should specify these numerous details when talking about security price.

We will take this observation a step further. Technically speaking, other than at the time of transaction we cannot say that price exists as a single number at all. Let us demonstrate this point. In financial markets securities are normally bought and sold in an exchange, and the record of current orders is called an order

book. An example of order book is shown in Fig. 1. The order book shows how buyers (on the left side) are willing to buy the security at a lower price and sellers (on the right side) are willing to sell it at a higher price[1]. The trading parties wait in line for a matching order, and until that order arrives, the security does not have a single price. Instead, there is a spectrum of prices, that could potentially represent the security price.

Best candidates among them are the best bid $s_{bid}$ and the best ask $s_{ask}$ (marked green and red in Fig. 1). The difference between the two is called the bid-ask spread:

$$\Delta = s_{ask} - s_{bid}, \tag{1}$$

We can say that the security price is localized between the best bid and the best ask. When an order is matched it disappears from order book and a transaction is recorded. Then the transaction price can be referred to as the security price, but again, within the context of a particular transaction.

| **Buyers – bid** | | **Sellers - ask** | |
|---|---|---|---|
| **Price** | **Size** | **Price** | **Size** |
| 27.83 | 100 | 27.87 | 100 |
| 27.82 | 100 | 27.9 | 100 |
| 27.8 | 200 | 27.95 | 1000 |
| 27.79 | 600 | 28.15 | 300 |
| 27.78 | 100 | 28.2 | 400 |

Fig. 1. Sample order book. Buyers (on the left side) want to buy the security at a lower price and sellers (on the right side) want to sell it at a higher price. Best bid is marked green, and best ask is marked red.

Often quotes in the order book come from market makers. Market makers are companies or desks within companies that quote buy and sell prices of financial instruments for other market participants, while providing commitment to buy and sell at the quoted prices. These firms profit from the bid-ask spread, and spread management is crucial for them. As market makers, they compete for order execution and larger turnover on the assets they quote.

[1] Sometimes there might be no buyers or sellers, or neither buyers, nor sellers

Higher turnover is easily achieved by reducing the operating spread – the difference between market maker's own buy and sell quotes. However, reducing the operating spread also lowers income per trading cycle[2]. To the contrary, increasing the operating spread increases income per cycle, but reduces turnover. There is a tradeoff between spread and turnover, and the fundamental question is: what is the optimal operating spread, that will maximize market maker's profit from a given security?

Understanding spread behavior is important not only for market makers, but also for market users and passive participants. It helps price discovery and allows them to save money by executing closer to the fair price. Companies managing larger funds, particularly pension funds, can use it to price larger blocks of securities. Understanding spread also provides tools for accurate risk management of securities with limited liquidity.

Spread has a deep fundamental value. In order to demonstrate this, let us try to answer a question: what is a measurement in finance and how is price measurement performed, [1-3]? If a number represents a valid security price, then there must be parties in the market willing to transact the security at that price. In order to test if the price is right, we must submit an order, for example a BUY order, at a discount price and keep increasing the price until somebody wishes to sell at our price. Once our order executes, we can say that price has been measured and the transaction price represents a valid security price. In fact, **every transaction in financial markets is an elementary act of price measurement.**

How do we improve price accuracy? If we start with a small order it may not have enough weight to represent price, so we may want to increase order size. It may work to a certain extent. However, at some point the order will become so large that it will affect price. Other traders will see it and adjust their orders, or our order will execute piercing multiple levels of order book – there are many ways how it can happen. Even though the spread may still be low, price itself will become distorted. Apparently, there is an inherent price uncertainty associated with the nature of price measurement. That uncertainty cannot be reduced and is directly related to spread.

This quality has been pointed out in [4-9] to resemble the Heisenberg's uncertainty principle in quantum mechanics. In our earlier work [2] we showed that spread can be described as a quantum notion with fluctuating coefficients, and that it obeys the statistics of quantum chaotic systems. In another work we showed that, not only spread, but entire price evolution can be described with quantum chaotic framework [1].

---

[2] Market maker's trading cycle consists of buying a security low from the current seller and selling it high to the next available buyer (or shorting and covering it).

The most common approach is to obtain spread as a result of modeling processes in the order book. Once order arrival, cancellation, and execution have been properly modeled, spread is obtained by direct calculation [10,11]. Another approach is to obtain spread as the optimal value from market maker's perspective [12,13]. This approach allows to calculate spread from microstructural parameters, but also depends on inventory held by market makers and their risk aversion level.

We are going to base our study on the considerations described in this section. Our goal is to express the theoretical concept of spread as a bandgap in a lwo-level system [3,14] through observable and measurable quantities. We are concerned with properties that have universal form, not depending on particulars of an exchange system or company fundamentals. Naturally for financial industry, we are not looking for 100% deterministic solutions. All relationships we find have statistical nature. But despite the statistical dispersion, certain behavioral characteristics can be factored out to be applied practically.

Unlike our previous publications, here we will differentiate between bid-ask spread and high-low bars. Bid-ask spread is related to the difference between the best bid and best offer, and is important to those who want to provide liquidity based on its demand. High-low bars are related to fixed timeframe and are important to those who simply want to update their quotes once in that timeframe while maintaining certain execution level. As we shall see, there are substantial differences in behavior of the two quantities and that is why we will differentiate them in this work.

Usually the designated market makers are subject to maintaining certain conditions, such as minimum quotation time, maximum allowed spread or minimum turnover. We will overlook these details and will focus on general framework. Firms and trading desks can include these additional conditions as they apply to them.

And lastly, in order to maintain focus we will only consider the equity asset class. Same concepts can be applied to other asset classes.

## 2. Basic relationship between spread, volatility and volume

### *Spread as price uncertainty*

Many factors play role in spread formation, such as price uncertainty, transaction costs, holding premium, etc [15-20]. Among them price uncertainty is the largest and most immediate. It is possible to evaluate the degree of that uncertainty for a stock, assuming as usual that its price $s$ follows a Gaussian random walk with volatility $\sigma$:

$$ds = s\sigma\, dz, \tag{2}$$

where $dz$ is a random variable obeying standard normal distribution. As price evolves, it will drift up or down until it hits bid or ask orders. This will lead to a transaction, ending price uncertainty, since price becomes determined as a result of transaction. We can therefore write the spread $\Delta$ as:

$$\Delta \approx 2s\sigma\sqrt{\tau}, \tag{3}$$

where $\tau$ is the average transaction time. We can write even more generally as

$$\Delta = \lambda s\sigma\sqrt{\tau} \tag{4}$$

to include the market makers' premium for holding inventory and for uncertainty of limit order execution. Coefficient $\lambda$ is dimensionless and its value depends on the market maker's risk aversion profile, quoting strategy, technical capabilities, and security profile. It may vary slowly with market conditions, but will maintain its order of magnitude since major dependencies of spread have already been factored out.

We can estimate $\tau$ from trading volume $V$ and average transaction size $n$ as

$$\tau = \frac{n}{V} \tag{5}$$

It is important that $\sigma$ and $V$ be related to the same time interval. Combining the two equations together, we have:

$$\Delta \approx \lambda s\sigma\sqrt{\frac{n}{V}} \tag{6}$$

This goes along with common sense, since it shows that spread must increase with volatility and decrease with volume. Before studying how Eq. (6) relates to market data, let us examine other ways to obtain spread. They will provide additional insight into how it forms.

*Spread as straddle premium*

Since market making strategies are looking to execute within a narrow range, they are essentially variations of bets on two conditions: asset price (a) being contained in an interval within the spread (b) within the average transaction time $\tau$. Theoretically, such bets could also be placed with European style options with expiration time equal to $\tau$. Straddle strategy is an example of such bet[3]. If the security price remains within the spread, then normally the received premium should cover payables at the exercise. Therefore, the range of positive $P/L$ should be equal to spread. This comparison is not exact and depends on the specifics of market maker's execution. Exact replication may involve exotic options and more than just two-legged strategy, let alone its practical realization. Let us estimate the spread from these considerations.

The range of positive $P/L$ for a straddle is:

$$\Delta = 2(c + p) \tag{7}$$

where $c$ and $p$ are the call and put premiums received. One can estimate these premiums using the approximation for ATM options resulting from the Black-Scholes model:

$$c_{ATM} = p_{ATM} \approx \frac{s\sigma\sqrt{\tau}}{\sqrt{2\pi}} \tag{8}$$

Combining all equations gives the final result:

$$\Delta \approx \sqrt{\frac{8}{\pi}}\, s\sigma\sqrt{\tau} \tag{9}$$

As we can see, this method produces the same result as Eq. (6) with implied value of $\lambda = \sqrt{\frac{8}{\pi}} \approx 1.6$.

*Dimensional considerations*

In fact, the same relationship (up to a coefficient) could be obtained by employing only dimensional considerations. Indeed, the only parameters that could define the spread are the stock price $s$ with dimension of $[money]$, volatility $\sigma$ with dimension of $\left[\frac{1}{\sqrt{time}}\right]$, trading volume $V$ with dimension of $\left[\frac{quantity}{time}\right]$, and

[3] Practical realization of such strategy is of course impossible, but this comparison serves as a good reference point for estimates.

average executing order size $n$ with dimension of $[quantity]$. The only combination in which they compose a spread with dimension of $[money]$ is

$$\Delta \sim s\left(\frac{n\sigma^2}{V}\right)^m \tag{10}$$

where $m$ is some power exponent. Realizing that under normal trading conditions spread must be proportional to volatility, we figure $m = \frac{1}{2}$:

$$\Delta = \lambda s \sqrt{\frac{n\sigma^2}{V}} \tag{11}$$

which is the same as Eq. (6). Market makers could set coefficient $\lambda$ to any value they like. Smaller $\lambda$ will ensure more execution and bigger turnover, but will also result in larger residual risk from induced inventory. Larger $\lambda$ will ensure rare execution and smaller residual risk. One way to find a reasonable value for coefficient $\lambda$ is from market data. A market maker can then calibrate his $\lambda$ relative to the market.

*Relation to market data*

Fig. 2 demonstrates Eq. (6) in action. It shows the scatterplot of $\Delta$ with $\lambda = 3.5$ vs. average bid-ask spread for a number of stocks on March 16, 2016. That value of $\lambda$ provides the best fit with the market data.

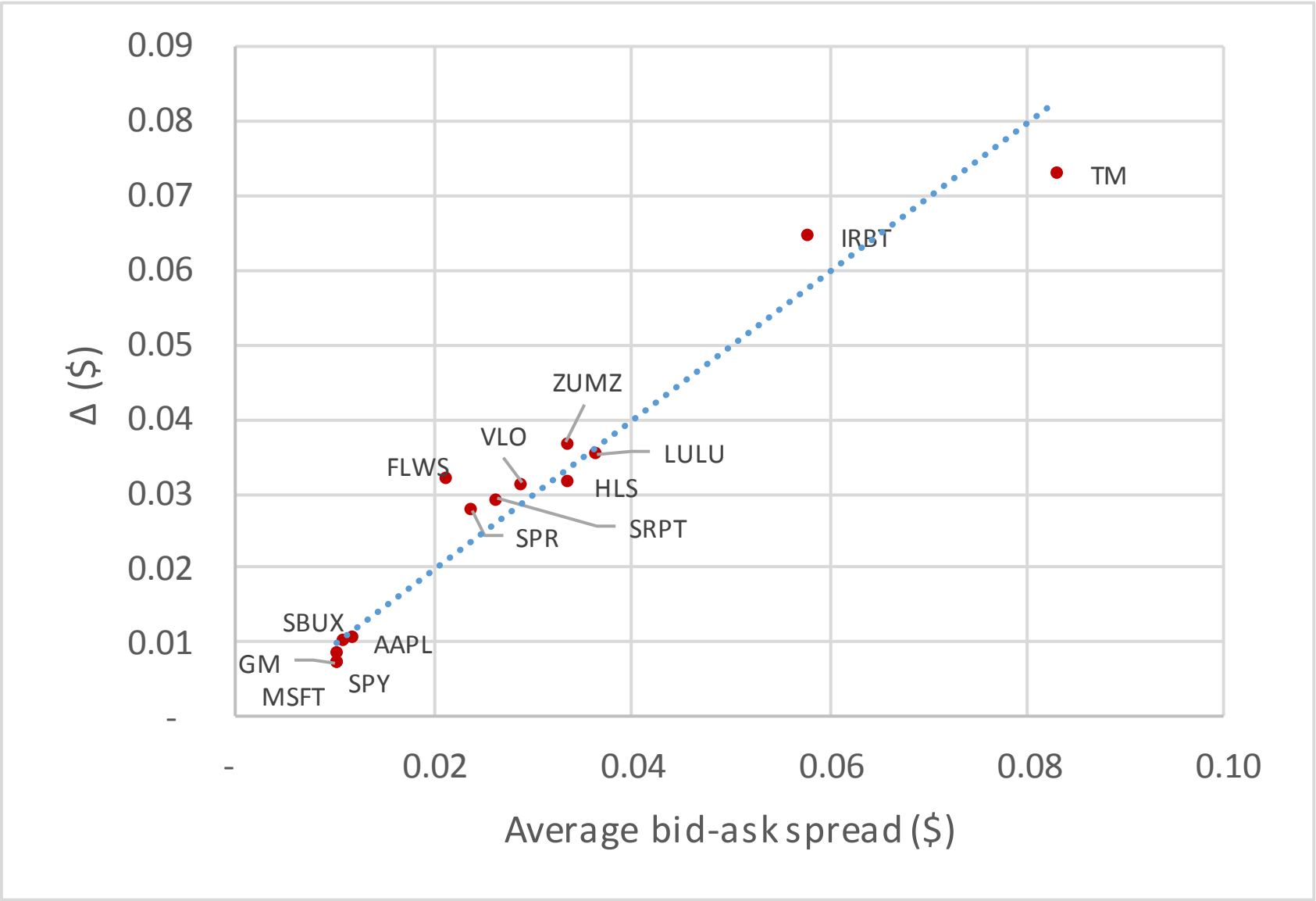


Fig. 2. Calculated spread $\Delta$ vs. average bid-ask spread on March 16, 2016.

Parameter $\lambda = 3.5$ has been used.

Fig. 3 shows how Eq. (6) works for spread dynamics. It displays calibrated Δ and average bid-ask spread in the period March 1-16, 2016 for AMZN (Amazon.com, Inc.) and LULU (Lululemon Athletica inc.).

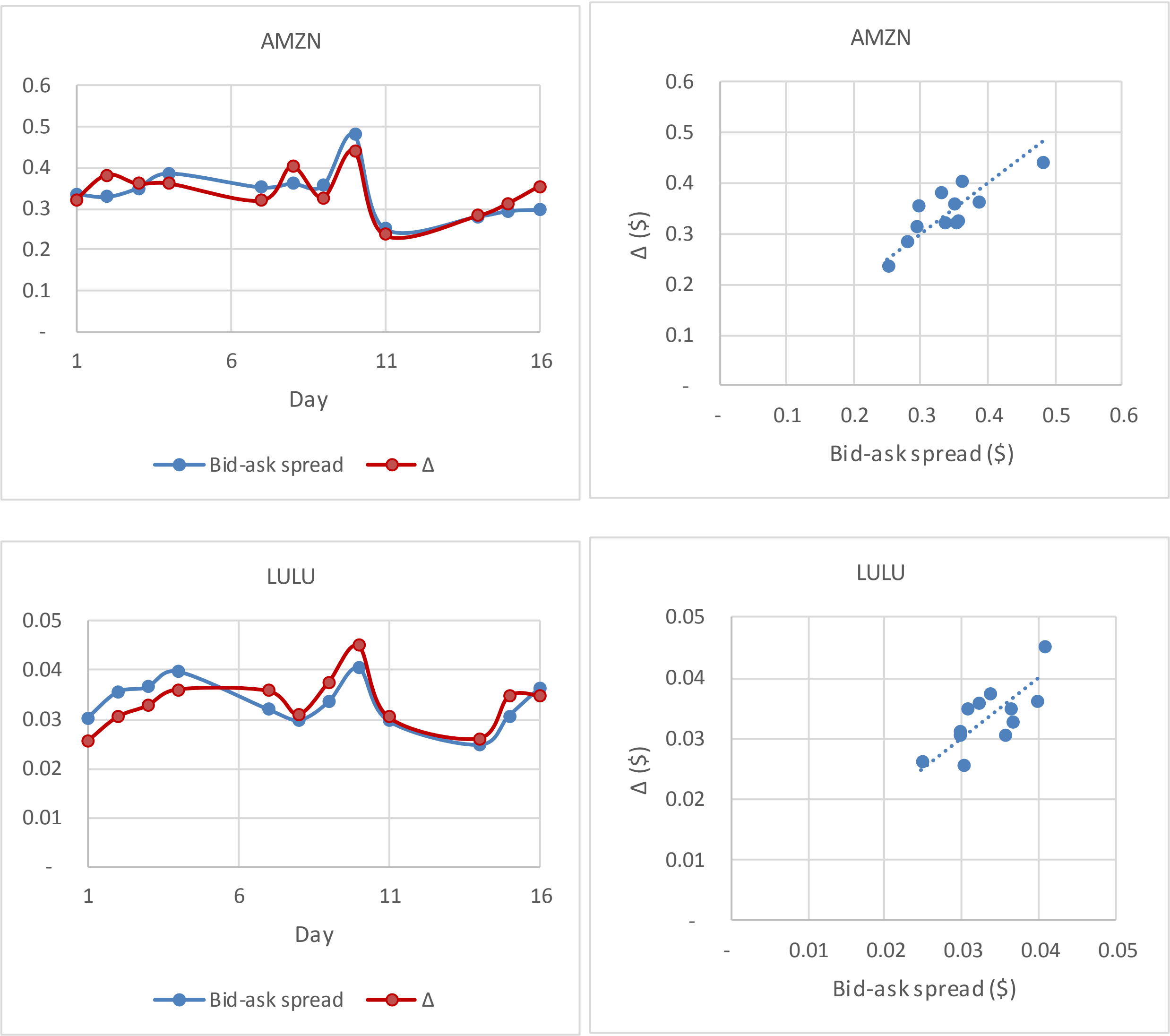


Fig. 3. Dynamics and correlation of calibrated Δ and average bid-ask spread

during the period of March 1-16, 2016 for AMZN and LULU.

Thus, we established that spread, volatility, and trading volume are not independent variables and are connected through Eq. (6). We also estimated the market's implied value for $\lambda$ to be around $\lambda = 3.5$. Relation Eq. (6) describes spread for various securities and captures its dynamics. It is important that no assumptions about the security, company, or market structure, other than that the security price follows a Wiener process, were made in deriving it.

## 3. Microscopic theory

*Overview of quantum coupled-wave model*

Eq. (6) says that spread must always decrease with volume. According to market data this is only true to a certain extent: when volume is large spread starts growing with volume, (see Figs. 11-13). In order to obtain the general relationship for spread, we need to use a model that deals with it as an intrinsic property of financial instruments and that captures its statistical properties.

Such model was developed in our earlier paper [2]. In that paper we proposed the idea that security prices can be described as eigenvalues of *price operator* acting on probability amplitude, so that:

$$\hat{S}\psi_n = s_n\psi_n \tag{12}$$

$$\text{and} \quad p_n = |\psi_n|^2 \tag{13}$$

Matrix elements of the price operator fluctuate in time, due to which its eigenvalues and eigenfunctions acquire random properties:

$$\hat{S}(t+\delta t) = \hat{S}(t) + \delta\hat{S}(t) \tag{14}$$

Here we will slightly modify it compared to [2] to build it for transfer between *high* and *low* levels, realizing that bid and ask are also a special case of high-low at the time scale equal to $\tau$. For a two-level system Eq. (12) takes the following form:

$$\psi = \begin{pmatrix} \psi_{high} \\ \psi_{low} \end{pmatrix} \quad \text{and} \quad \begin{pmatrix} s_{11} & s_{12} \\ s_{12}^* & s_{22} \end{pmatrix} \begin{pmatrix} \psi_{high} \\ \psi_{low} \end{pmatrix} = s_{high/low} \begin{pmatrix} \psi_{high} \\ \psi_{low} \end{pmatrix} \tag{15}$$

Given this, the *high* and *low* prices can be obtained as

$$s_{high} = s_{mid} + \frac{h}{2} \quad \text{and} \quad s_{low} = s_{mid} - \frac{h}{2} \tag{16a}$$

$$s_{mid} = \frac{s_{11}+s_{22}}{2} = \frac{s_{low}+s_{high}}{2} \quad (\text{mid} - \text{price}) \tag{16b}$$

$$h = \sqrt{(s_{11}-s_{22})^2 + 4|s_{12}|^2} \quad (\text{bar height or spread}) \tag{16c}$$

Matrix elements $s_{ik}$ of price operator are then parametrized to include fluctuations:

$$s_{11}(t+dt) = s_{last}(t) + s_{last}(t)\,\sigma dz + \frac{\xi}{2} \tag{17a}$$

$$s_{22}(t+dt) = s_{last}(t) + s_{last}(t)\,\sigma dz - \frac{\xi}{2} \tag{17b}$$

$$s_{12}(t+dt) = \frac{\kappa}{2} \tag{17c}$$

where $dz \sim N(0,1)$, $\xi \sim N(\xi_0, \xi_1)$ and $\kappa \sim N(\kappa_0, \kappa_1)$. In such setup the mid-price, the bar height and last price are given by equations

$$s_{mid}(t+dt) = s_{last}(t) + s_{last}(t)\,\sigma dz \tag{18a}$$

$$h = \sqrt{\xi^2 + \kappa^2} \tag{18b}$$

$$s_{last}(t+dt) \sim unif(s_{low}, s_{high}) \tag{18c}$$

At each step the next mid-price takes a value normally distributed around the last price. The high and low levels take values $h/2$ above and below the mid-price. Then the next last price takes a random value uniformly distributed between the high and low levels. Price chart that can be generated by the model, Eqs. (18a-c) is shown in Fig. 4. Volatility of prices (as referring to last prices) is equal

$$\eta = \sqrt{s^2\sigma^2 dt + \alpha \frac{h^2}{4}} \tag{19}$$

where $\alpha$ is a coefficient that corresponds to distribution of $s_{last}$ in Eq. (18c)[4], and we used notation $s$ for $s_{last}$. Rather than considering bars as rigid boundaries beyond which price cannot extend, it is possible to think of them as the characteristic distribution width of the last price around mid-price. For example, that distribution can be normal with half the bar height as its standard deviation:

$$s_{last}(t+dt) \sim N\left(s_{mid}, \frac{h}{2}\right) \tag{20}$$

In this case volatility is:

[4] Statistics of trades inside the high-low range can be different for different time scales, which will result in different $\alpha$. For example, trades execute at bid or ask levels but not in the middle, but as we move up in time horizon, trades become more frequent in the middle of the high-low range and scarce at its edges.

$$\eta = \sqrt{s^2\sigma^2 dt + \frac{h^2}{4}} \tag{21}$$

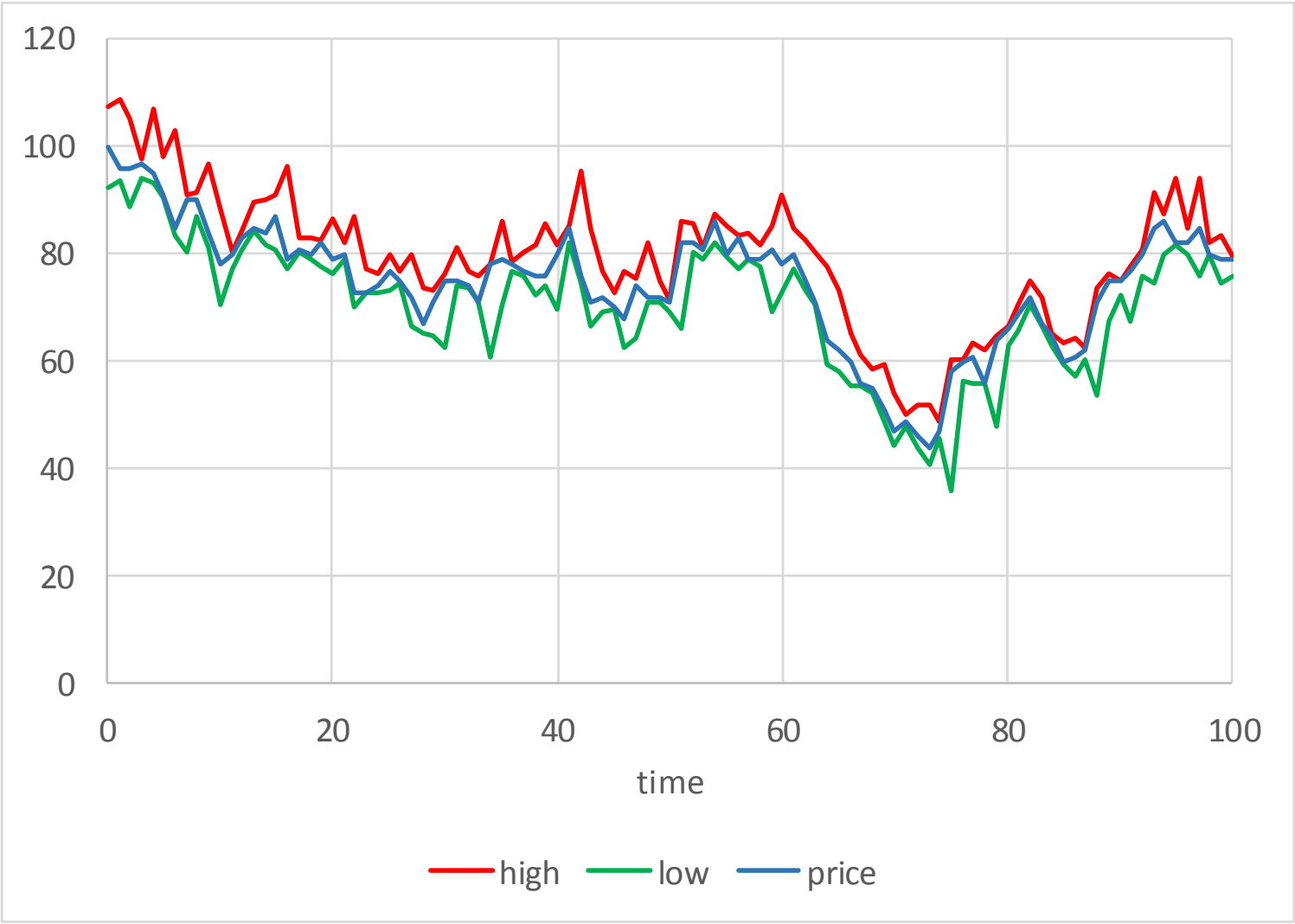


Fig. 4. Price chart generated by coupled-wave model. At each step the next mid-price takes a value normally distributed around the last price. The high and low levels take equally spaced values above and below the mid-price. Then the next last price takes a random value uniformly distributed between the high and low levels.

Bars described by Eq. (18b) behave as quantum-chaotic quantities [2], whose statistics matches the observed statistics quite well both on bid-ask micro-level and on bar data level, see Figs. 5a, 5b.

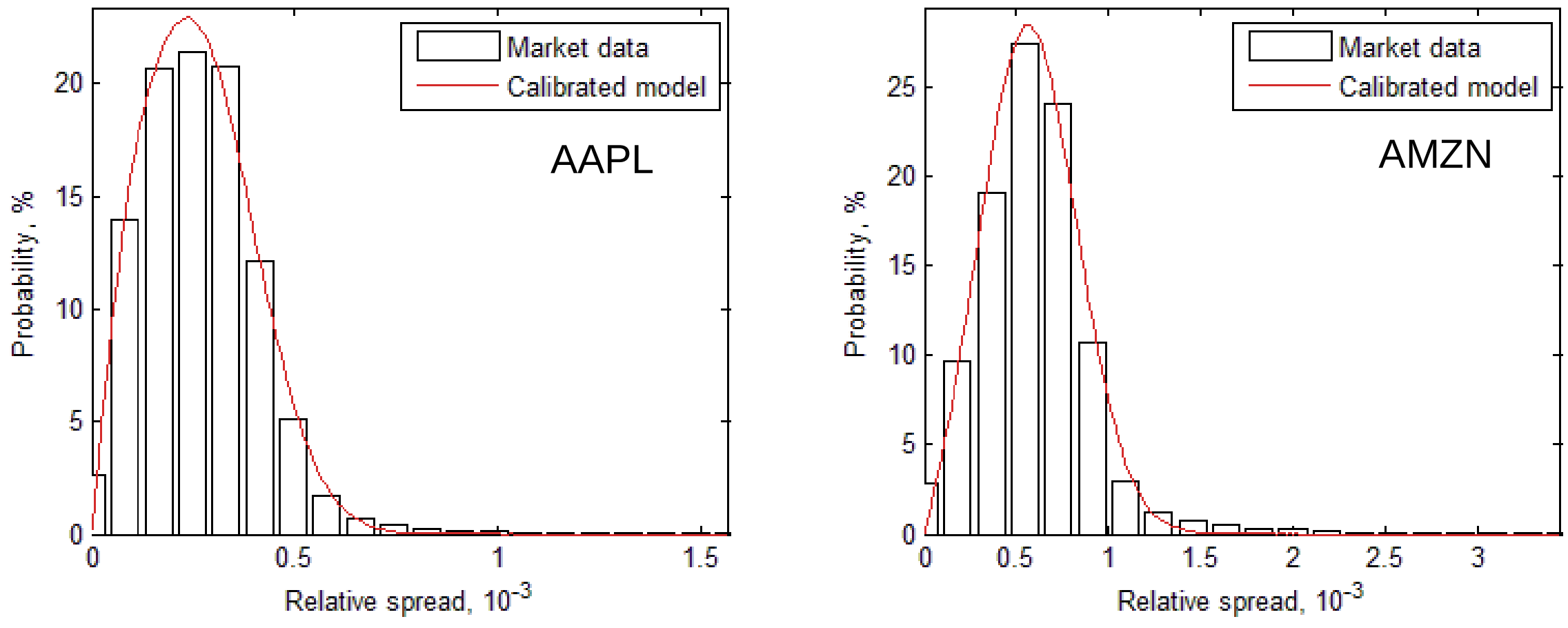


Fig. 5a. Calibration of coupled-wave model to bid-ask data for AAPL and AMZN, [2]

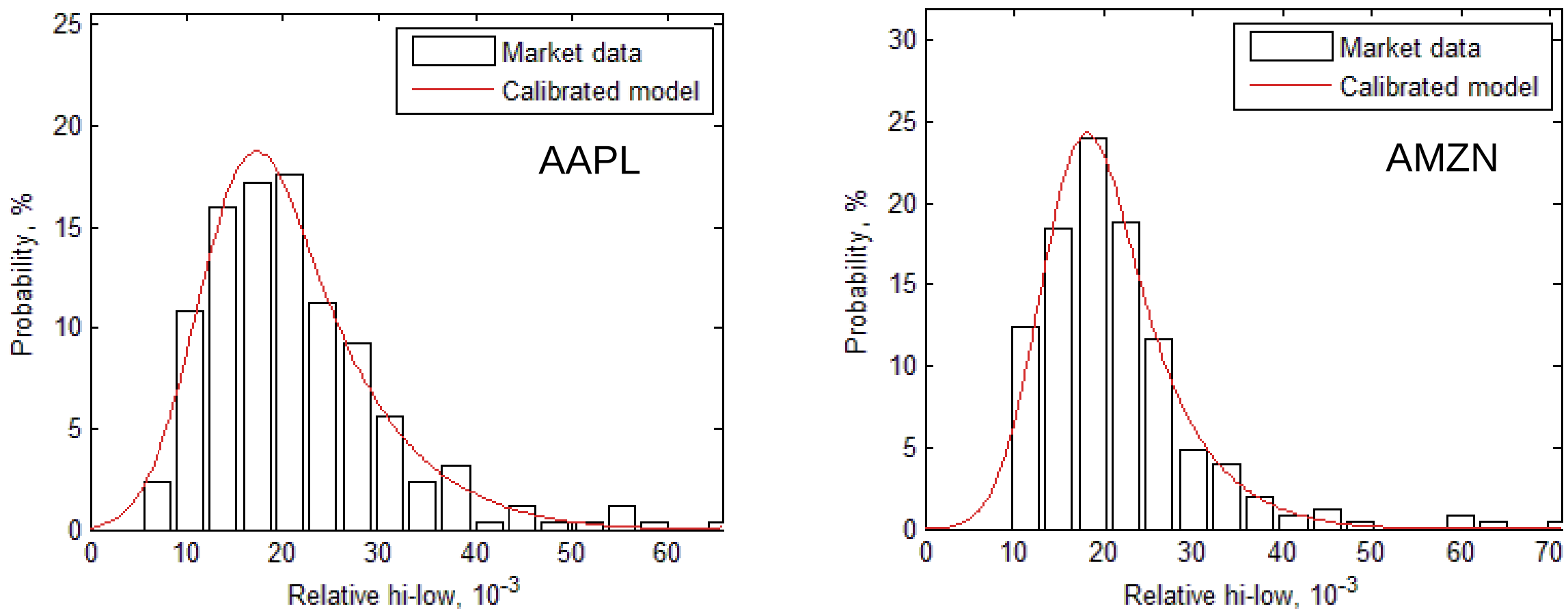


Fig 5b. Calibration of coupled-wave model to bar data for AAPL and AMZN, [2]

Evolution of probability amplitude $\psi$ is described by equation [1,2]:

$$i\tau_0 s \frac{\partial \psi}{\partial t} = \hat{S}\psi \tag{22}$$

or in an open form

$$i\tau_0 s \frac{d\psi_{high}}{dt} = s_{11}\psi_{high} + s_{12}\psi_{low} \tag{23a}$$

$$i\tau_0 s \frac{d\psi_{low}}{dt} = s_{12}^*\psi_{high} + s_{22}\psi_{low} \tag{23b}$$

Here $\tau_0$ is some constant with dimension of time. For constant coefficients this system of equations has the following solution, expressed through model parameters $h$, $\xi$, and $\kappa$ [2]:

$$\psi_{high}(t) = e^{-is_{mid}t}\left\{\left[cos\left(\frac{h}{2\tau_0 s}t\right) - i\frac{\xi}{h}sin\left(\frac{h}{2\tau_0 s}t\right)\right]\psi_{high}(0) - i\frac{\kappa}{h}sin\left(\frac{h}{2\tau_0 s}t\right)\psi_{low}(0)\right\} \tag{24a}$$

$$\psi_{low}(t) = e^{-is_{mid}t}\left\{-i\frac{\kappa}{h}sin\left(\frac{h}{2\tau_0 s}t\right)\psi_{high}(0) + \left[cos\left(\frac{h}{2\tau_0 s}t\right) + i\frac{\xi}{h}sin\left(\frac{h}{2\tau_0 s}t\right)\right]\psi_{low}(0)\right\} \tag{24b}$$

Since in fact coefficients $s_{ij}$ fluctuate, solution has to be applied numerically in small time steps, during which price operator elements $s_{ij}$ can be considered constant.

*General relation for spread*

Coupled-wave model contains three dimensions of uncertainty: (a) uncertainty of the mid-price, Eq. (18a), (b) uncertainty of bar size, Eq. (18b), and uncertainty of price within the bar, Eq. (18c). The first two elements accumulate over time, while the uncertainty of price within the bar is present from the beginning. Spread must cover these risks over the liquidation period $\tau$. Scaling these risk components each according to its risk-aversion level, we can compose the spread to be

$$\Delta = \sqrt{\frac{(\rho h)^2}{4} + (\lambda\eta)^2} \tag{25}$$

where in this case $\eta = \sqrt{s^2\sigma^2\tau + \frac{\rho^2}{\lambda^2}\frac{h^2}{4}}$. The initial uncertainty is $\frac{h}{2}$ up and down, and over time $\tau$, it accrues $s\sigma\sqrt{\tau}$ and another half-bar $\frac{h}{2}$. To be accurate, we should have used different $h$ for the initial and subsequent bars, but for our purposes we will be using characteristic values. Eventually, we have

$$\Delta = \sqrt{\frac{(\rho h)^2}{2} + (\lambda s\sigma)^2\tau} \tag{26}$$

Let us link all components to observable quantities. The second term under the square root is the already familiar element associated with price uncertainty due to finite liquidation time. It is essentially the price that a buy-side trader must pay in order to access immediate liquidity. We can call it the *liquidity price*.

Bar size $h$ determines oscillation frequency in Eqs. (24a,b). Oscillation period corresponds to double the average transaction time $2\tau$, in which the security is transferred back and forth in a full cycle. We must therefore have

$$\frac{h\tau}{2s\tau_0} \approx \pi \tag{27}$$

and as a result:

$$h \approx 2\pi\tau_0\frac{s}{\tau} \tag{28}$$

Noting that $\frac{s}{\tau}$ is the amount of money per share traded in a transaction, we come to conclusion that $h$ represents money flow and characterizes the degree of price impact caused by that flow. We will call it the *impact price*.

Within $h$, parameter $\kappa$ is associated with securities transfer between the "high" and "low" levels, and parameter $\xi$ is associated with the intensity of that transfer. This can be verified by direct modeling of Eqs. (24a,b) and varying $\xi$ and $\kappa$, particularly using combinations $\xi = 0, \kappa \neq 0$ and $\kappa = 0, \xi \neq 0$.

Combining Eqs. (5, 26, 28), we have the final result for the spread:

$$\Delta = \sqrt{\left(\lambda s\sigma\sqrt{\tau}\right)^2 + 2\left(\rho\frac{\pi s\tau_0}{\tau}\right)^2} = \sqrt{\lambda^2\frac{s^2\sigma^2 n}{V} + 2\rho^2\left(\frac{\pi s\tau_0}{n}\right)^2 V^2} \tag{29}$$

We can shape up this equation writing it in a simpler dimensionless format:

$$\delta(v) = \sqrt{\frac{a}{v} + v^2} \tag{30}$$

where $\delta = \frac{\Delta}{s}$, $a = \sqrt{2}\rho\lambda^2\sigma^2(\pi\tau_0)$, $v = \frac{V}{V_0}$, and $V_0 = \frac{1}{\sqrt{2}\rho}\frac{n}{\pi\tau_0}$.

Eq. (30) relates the spread to microstructural parameters and is more general than Eq. (6). We see that spread can have two regimes with different characteristic behavior. When volume is small, such that $v \ll \sqrt[3]{a}$, the liquidity price contribution prevails over impact price contribution and spread exhibits the already familiar behavior: $\delta(v) \sim \frac{1}{\sqrt{v}}$. This regime is shown in Fig. 6a, where we can see how adding more flow to trading reduces the spread and helps improve price. This is valid as long as volume is not large enough to affect price. When $v \gg \sqrt[3]{a}$, so that cash value of executing orders becomes larger than the liquidity price, these orders begin to impair price measurement and spread starts to grow linearly with volume: $\delta(v) \sim v$. This regime is shown in Fig. 6b.

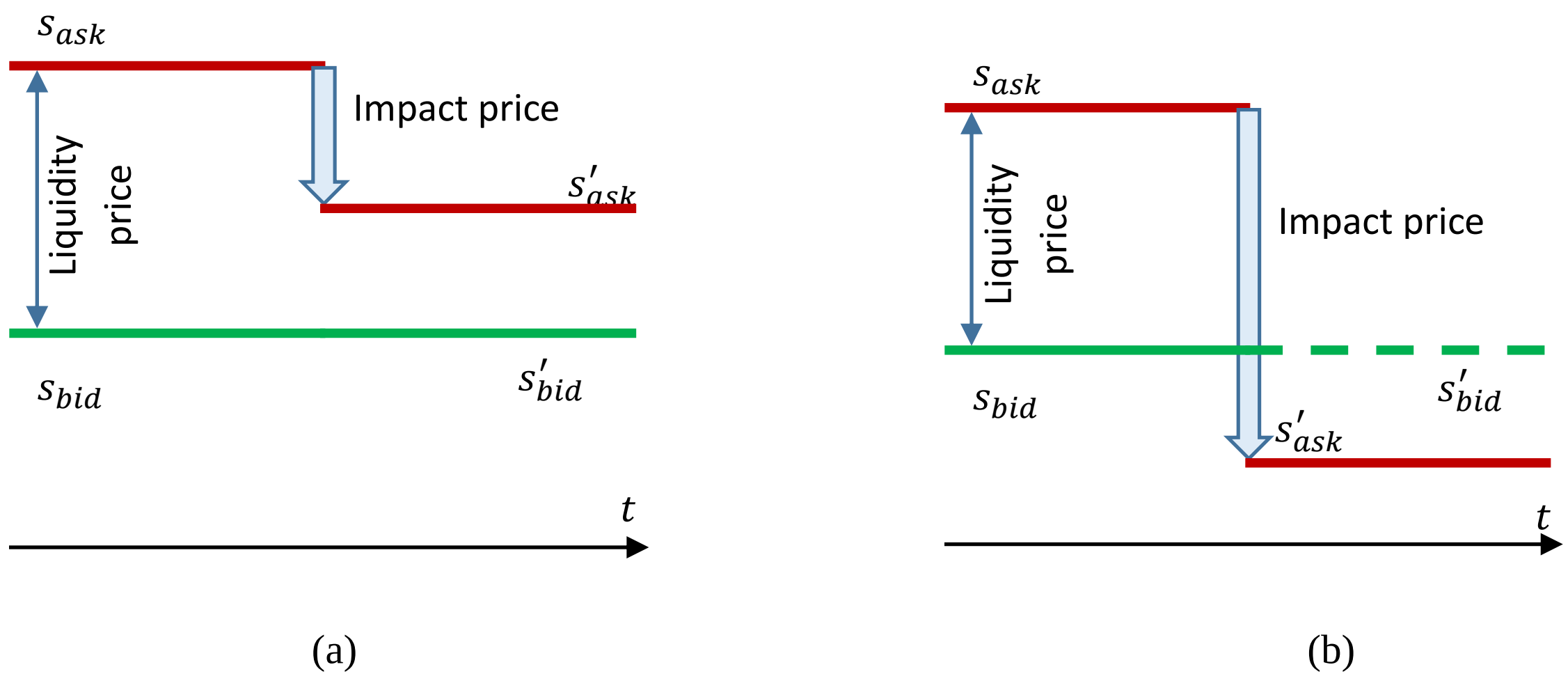


Fig. 6. Interplay between liquidity price and impact price: (a) impact is small and adding more liquidity improves price accuracy, (b) impact is so large that it impairs price.

*Minimum spread and price uncertainty*

Due to the functional form of Eq. (30), $\delta(v)$ does not reach zero. It reaches minimum at $v_{min} = \sqrt[3]{\frac{a}{2}}$, and its minimum value is equal to $\delta_{min} = \sqrt{3}\, v_{min}$. For every spread $\delta > \delta_{min}$ there are two values of $v$ that correspond to it.

*Time scaling of spread*

What happens if a desk quotes prices based on some time scale and wants to change to another time scale[5]. What is the relationship between high-low bars at different time scales? How do we transition from bid-ask spread to bars, or between the bars of different time scales?

If we were to quote at a different time scale with the same risk aversion, we would just use the volatility related to that time scale in Eq. (25) and add it to the initial uncertainty, represented by the initial bar:

$$\Delta_T = \sqrt{\frac{\rho^2 h_\tau^2}{4} + \lambda^2 \eta_\tau^2 \frac{T}{\tau}} \tag{31}$$

where indexes now indicate the reference time, and $T \geq \tau$. Expressing $\Delta_{T_2}$ through the parameters related to reference time $T_1$, we get

$$\Delta_{T_2} = \Delta_{T_1} \sqrt{1 + \lambda^2 \frac{\eta_{T_1}^2}{\Delta_{T_1}^2}\left(\frac{T_2}{T_1} - 1\right)} \tag{32}$$

Particularly, $\Delta_T$ is related to bid-ask spread $\Delta_\tau = \sqrt{\frac{\rho^2 h_\tau^2}{4} + \lambda^2 \eta_\tau^2}$ through:

$$\Delta_T = \Delta_\tau \sqrt{1 + \lambda^2 \frac{\eta_\tau^2}{\Delta_\tau^2}\left(\frac{T}{\tau} - 1\right)} \tag{33}$$

This agrees with the result obtained in our other work [1] (Eq. (43) in it). Equations are mapped to each other with the following substitutions:

---

[5] Some desks do tick-based quoting, which is tied to market events, while others do bar-based quoting refreshing their quotes after certain time. But even the bar-based quoters require this transition since they normally randomize their time horizon.

$$\Delta_\tau \leftrightarrow w_{\Delta \mathrm{t}}$$

$$\lambda\eta_\tau \leftrightarrow \beta\epsilon$$

$$\tau \leftrightarrow \Delta t$$

Some spread curves calibrated to average intraday minute bars, observed on March 16, 2016, and average daily bars are shown in Figs. (7a and 7b).

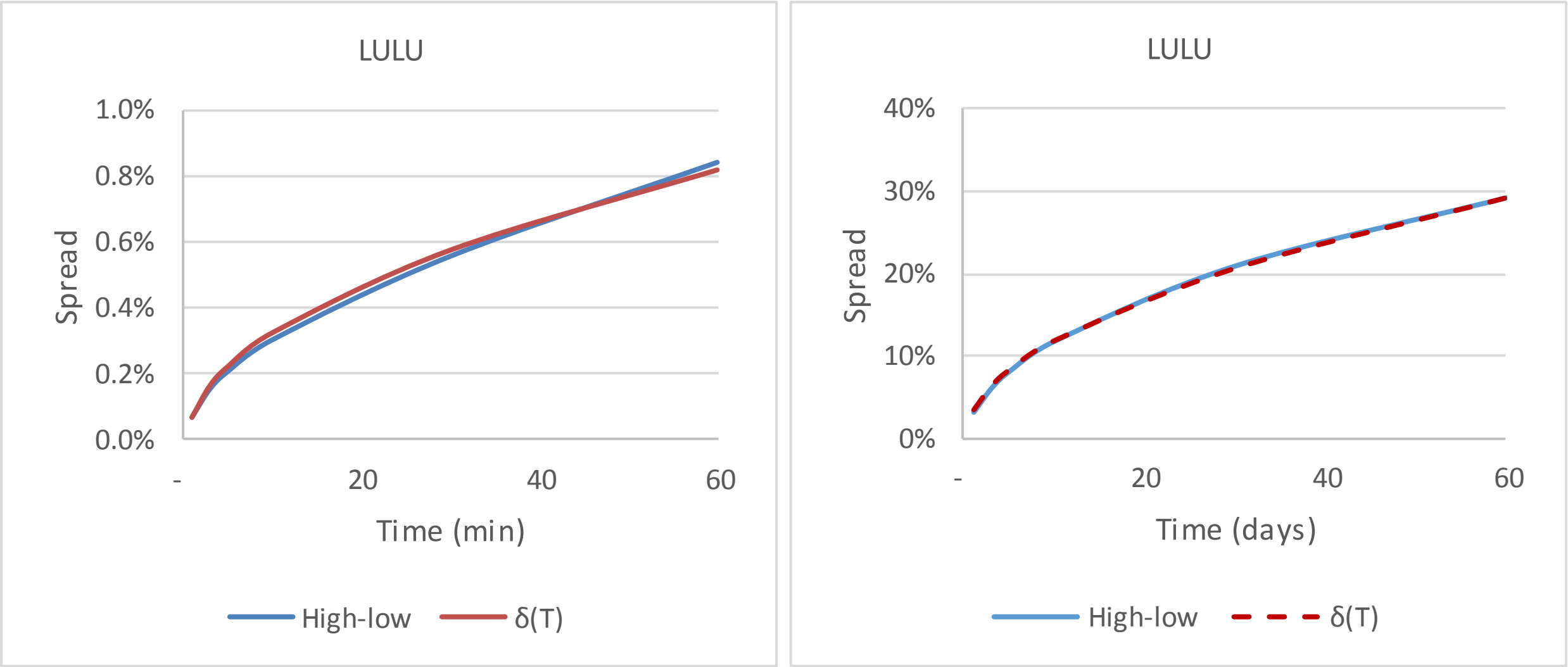


Fig. 7a. Spread curves calibrated to average 1-minute intraday (left) and daily (right) bars for ticker LULU

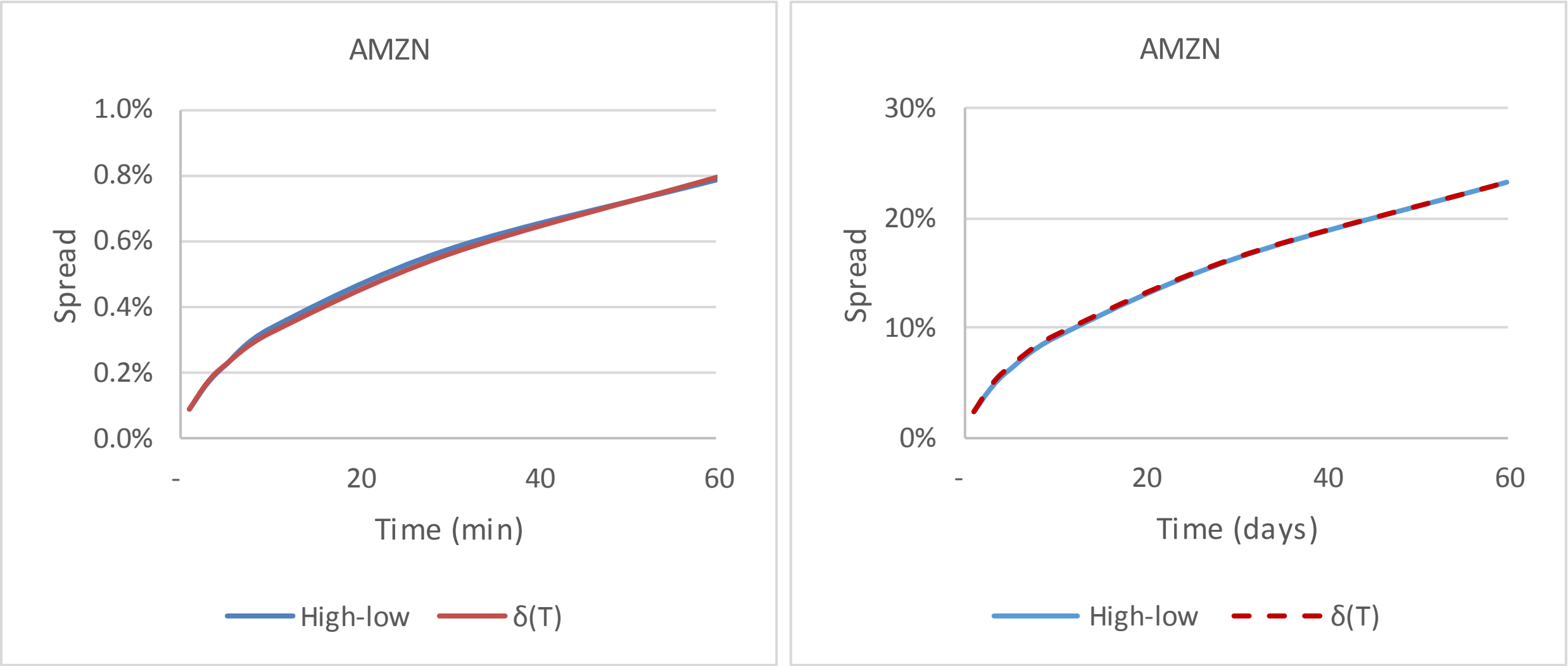


Fig. 7b. Spread curves calibrated to 1-minute intraday (left) and daily (right) bars for ticker AMZN

*Comparison to classical theory*

According to classical theory volatility scales as square root of time[6]:

$$\eta_{T_2} = \sqrt{\frac{T_2}{T_1}}\eta_{T_1} \tag{34}$$

This implies that price accuracy can be indefinitely improved by reducing measurement time. This is not true in real markets, as was discussed in Introduction section, and Eq. (33) does not allow it. Only over long time horizon volatility prevails and the effect of the initial spread disappears, leading to regular square root-like behavior:

$$\Delta_{T_2} \approx \sqrt{\frac{T_2}{T_1}}\Delta_{T_1} \tag{35}$$

To see the difference visually, we can take the spread curve of the LULU stock. If a 1-minute high-low bar was scaled forward to 60 minutes with Eq. (35), the result would be 0.55%, which about 35% off the real value.

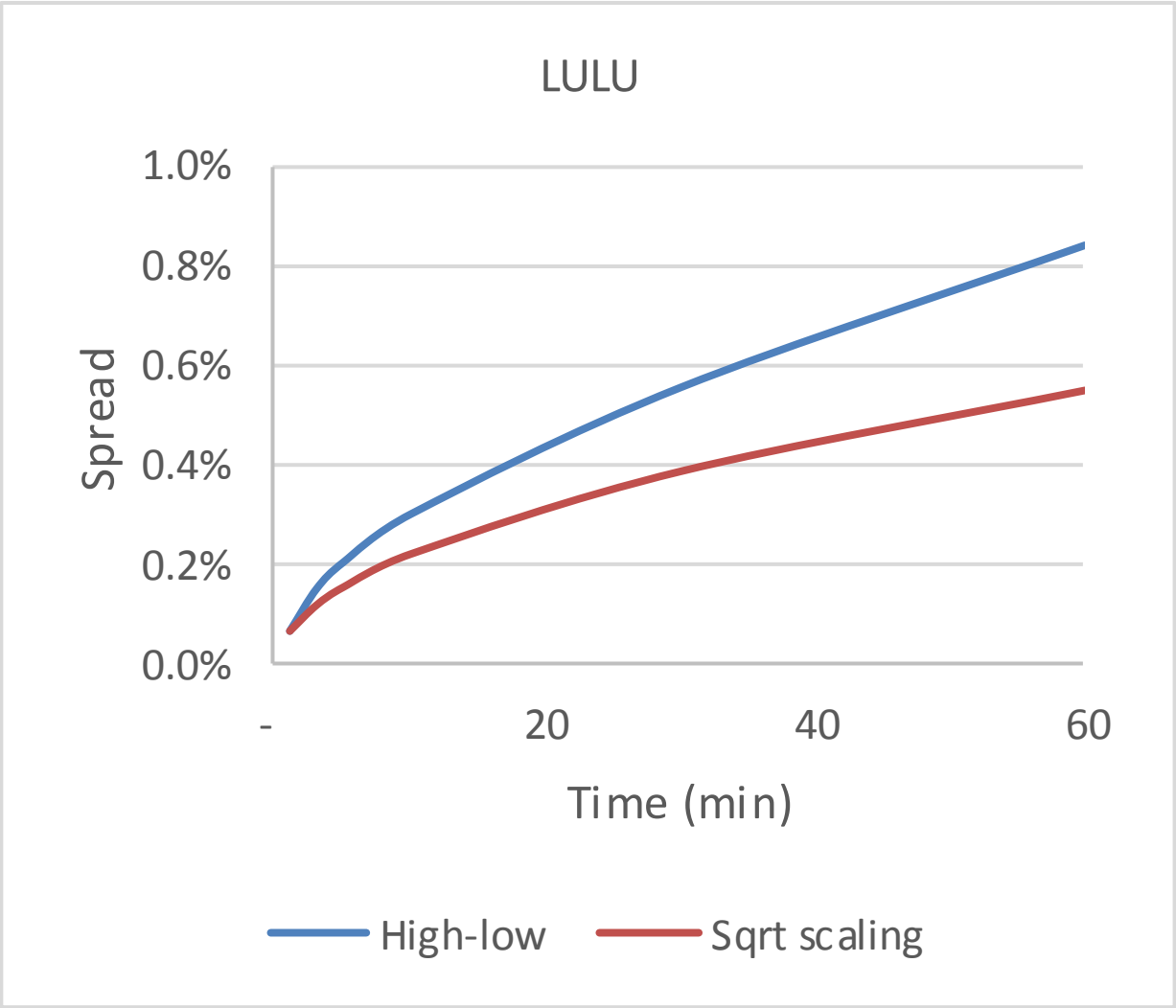


Fig. 8. Comparison of "quantum" spread curve, obtained with Eq. (32) and classical spread curve, obtained with the square-root of time scaling.

[6] Strictly speaking, this relates to volatility of returns. However, since return $r = \frac{s}{s_0} - 1$, the relation is the same for small price deviations discussed here.

*Scaling law including volume*

Eq. (32) can be made to include variation of spread with volume. Beginning with Eq. (25) and transitioning from bid-ask spread to a fixed bar time $T$, we get:

$$\Delta_T = \sqrt{\rho^2\left(\frac{\pi s\tau_0}{\tau}\right)^2 + \left[\lambda^2\left(s\sigma_\tau\sqrt{\tau}\right)^2 + \rho^2\left(\frac{\pi s\tau_0}{\tau}\right)^2\right]\frac{T}{\tau}} = \\ = \sqrt{\lambda^2\left(s\sigma_\tau\sqrt{T}\right)^2 + \rho^2\left(\frac{\pi s\tau_0}{\tau}\right)^2\left(1+\frac{T}{\tau}\right)} \tag{36}$$

Expressing it through volume, we get:

$$\Delta_T = s\cdot\sqrt{\lambda^2\sigma_\tau^2 T + \rho^2\left(\frac{\pi\tau_0}{n}\right)^2 V^2\left(1+\frac{VT}{n}\right)} \tag{37}$$

Noting that $\sigma_\tau^2 T = \sigma_T^2 1_T$, where $\sigma_T$ is simply the volatility of the mid-price variations of high-low bars and $1_T$ is a unity with dimensions of $T$, we have

$$\Delta_T(V) = s\cdot\sqrt{\lambda^2\sigma_T^2 1_T + \rho^2\left(\frac{\pi\tau_0}{n}\right)^2 V^2 + \rho^2\frac{(\pi\tau_0)^2 T}{n^3}V^3} \tag{38}$$

or in dimensionless format:

$$\delta_T(v) = \sqrt{\lambda^2\sigma_T^2 1_T + \frac{v^2}{2} + \frac{T}{2^{3/2}\rho\pi\tau_0}v^3} \tag{39}$$

We see that high-low bars depend on volume completely differently than bid-ask spread. Unlike $\delta_\tau$ the $\sigma$ term for $\delta_T$ does not depend on volume, and along with $v$ there is now a $v^{3/2}$ behavior, which prevails at large time horizons. Unlike $\delta_\tau$, which has a minimum, $\delta_T$ starts with its minimum value at $v_T = 0$ and only increases with volume. The differences can be seen in Fig. (9) showing characteristic $\delta_\tau(v)$ and $\delta_T(v)$ behavior.

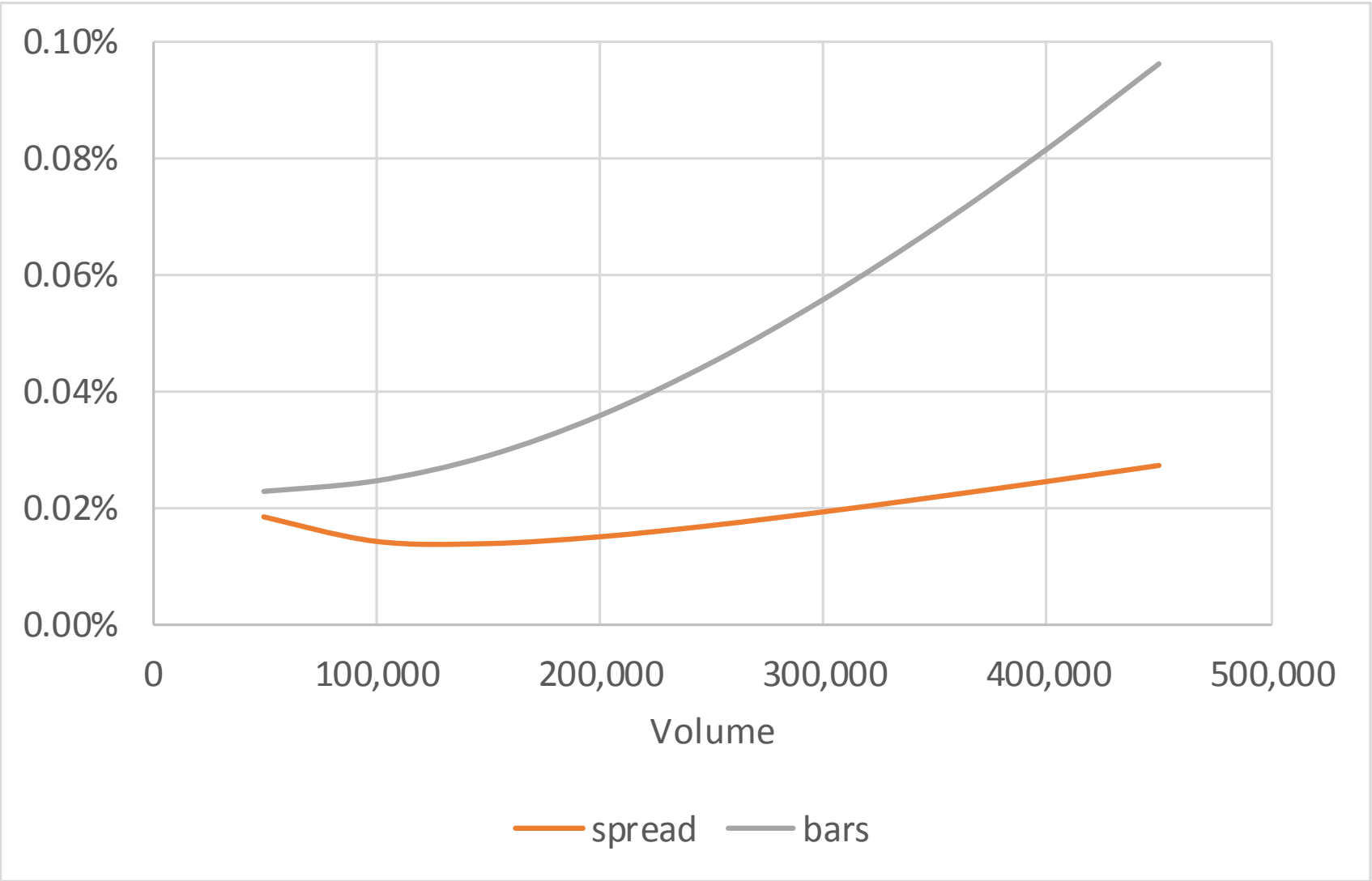


Fig. 9. Qualitative difference in behavior between (a) bid-ask spread and (b) high-low bars.

*Risk aversion level*

Eqs. (29 and 38) provide a good reference, but not yet the final answers to spread modeling. One reason is that statistics of trades inside the high-low range can be different for data at different time scales, which will result in different $\alpha$ in Eq. (19). For example, trades execute at bid or ask levels, but not inside the spread. Same trades can be more evenly distributed between the high and low levels on minute time scale. Lastly, in daily data trades usually occur around the center of high-low interval rather than by its edges. Transitions between time scales must include variation of $\alpha$ with time scale.

Another point is that risk aversion level depends on numerous factors. For example, market makers would usually ask a bigger premium to quote at smaller time scale than at large time scale. Additionally, they will generally set larger spreads at market opening since they are unsure about market consensus regarding prices. They will subsequently lower spreads once such consensus is established. Market makers will also increase spreads closer to the end of the day in order to prepare for longer holding timeframe (until market open) or to reduce chances of carrying overnight positions.

As complex as they might be, these details are ultimately just various forms of risk. They affect coefficients $\lambda$ and $\rho$, making them dependable on $T$ and $V$. Taking this into account, we can write the adjusted equations:

$$\delta_\tau = \sqrt{\lambda_{\tau,V}^2 \frac{\sigma_\tau^2 n}{V} + 2\rho_{\tau,V}^2 \left(\frac{\pi\tau_0}{n}\right)^2 V^2} \tag{40}$$

$$\delta_T(V) = \sqrt{\lambda_{T,V}^2 \sigma_T^2 1_T + \rho_{T,V}^2 \left(\frac{\pi\tau_0}{n}\right)^2 V^2 + \rho_{T,V}^2 \frac{(\pi\tau_0)^2 T}{n^3} V^3} \tag{41}$$

Despite that, variations of $\lambda$ and $\rho$ are small, since the most important variation parameters have already been factored out.

Bar height $\delta_T$ as a function of $v$ and $T$ is represented by a surface, that can be used to calibrate the model to market data.

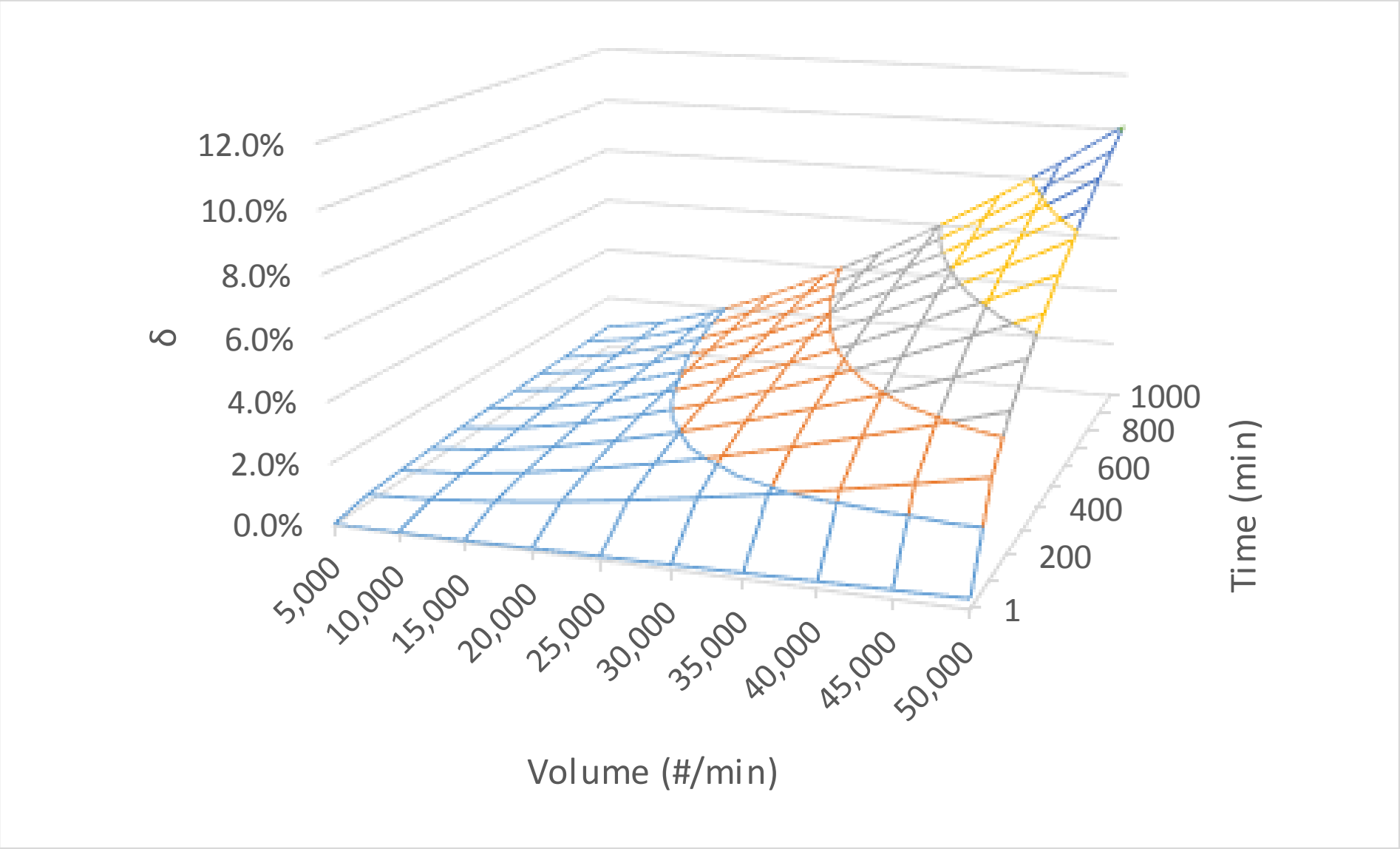


Fig. 10. Spread surface for AMZN.

It is natural to use $\lambda$ and $\rho$ as control parameters allowing to gauge execution rate. Numerical connection between the two will be established in the next section.

*Connection with market data*

As an example Figs. 11-13 present spread-volume data based on bid-ask spread, 1-minute intraday high-low bars, and daily high-low bars for AMZN and LULU. The spread-volume curves correspond to execution rate levels of 90%. They were obtained by taking the scatterplot of spread vs. volume, splitting it into volume buckets, and calculating the 90-th percentile across the spread for each bucket. Values for $n$

and $\sigma_\tau$ were measured directly from trading flow data. What was left after that is only to select $\tau_0$ and calibrate $\lambda$ and $\rho$. In order to make judgement of statistical significance easier, we supplement main plots with trade frequency using an additional vertical axis.

The model works pretty well for daily high-low data. It gets more complex for bid-ask and 1-minute high-low data. One can notice how the model works better on LULU than AMZN. This is because coupled-wave model assumes only two price levels at each step, and LULU has much more distinct levels than AMZN. This is apparent from trade frequency data, so LULU is better approximated by a two-level system. In order to describe AMZN more accurately, a multi-level model has to be used.

We can also see that 1-minute high-low bars do not start with a finite value, but grow from almost zero values (daily bars are fine in that respect). This is where the already mentioned effect of change in trade statistics from the spread to bars comes into play. Best bid and ask prices are achieved more easily than values beyond them. Trades inside the spread have an M-shaped distribution vs. smoother statistics in 1-minute bars.

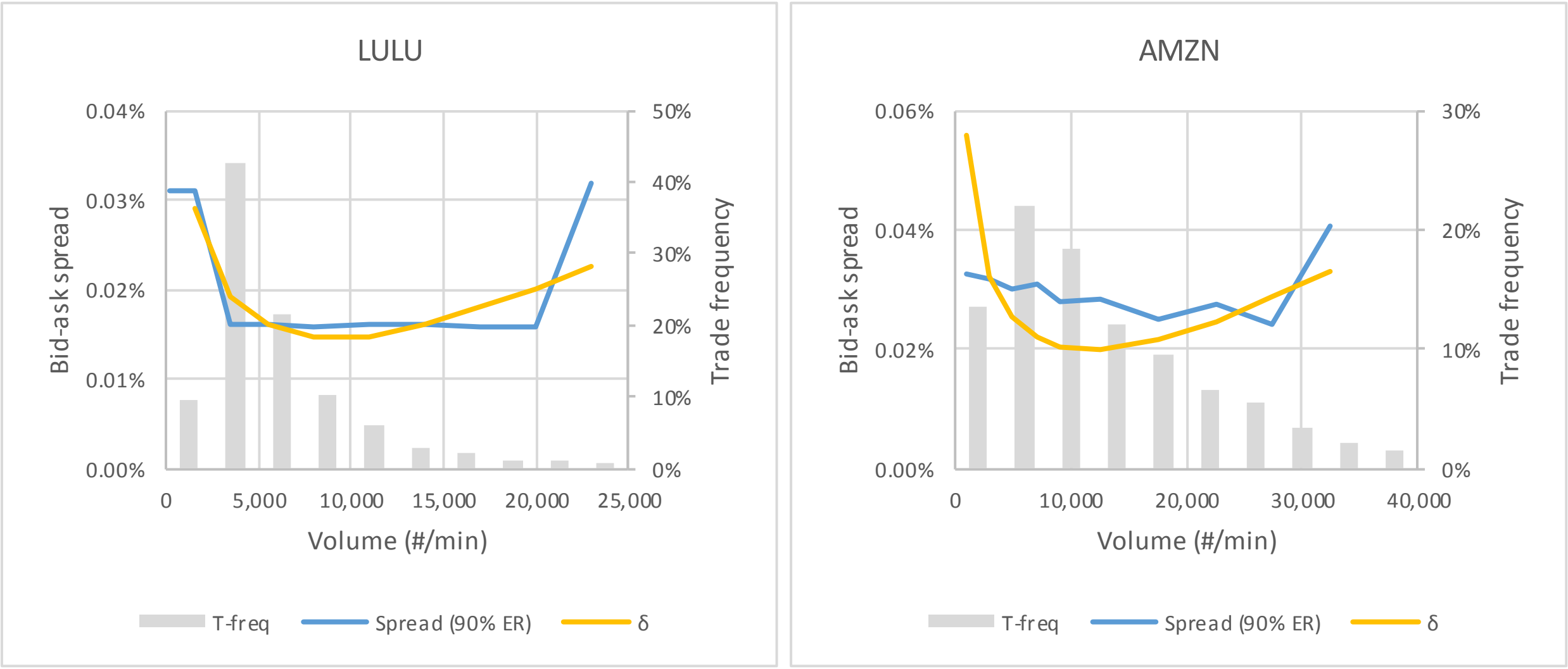


Fig. 11. Spread-volume curves based on bid-ask spread for LULU and AMZN. Lines correspond to execution rate (ER) 90%. Histogram represents the frequency of trades.

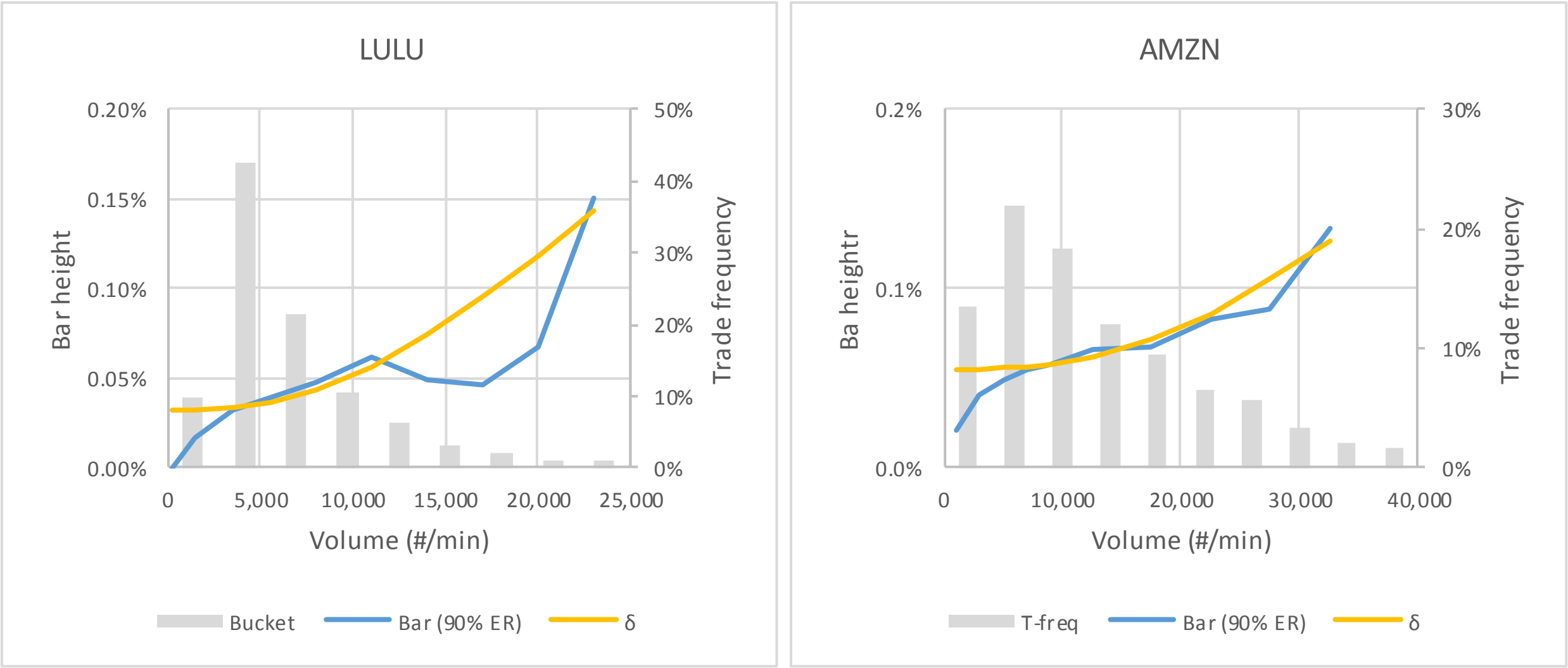


Fig. 12. Spread-volume curves based on 1-minute high-low bars for LULU and AMZN. Lines correspond to execution rate (ER) 90%. Histogram represents the frequency of trades.

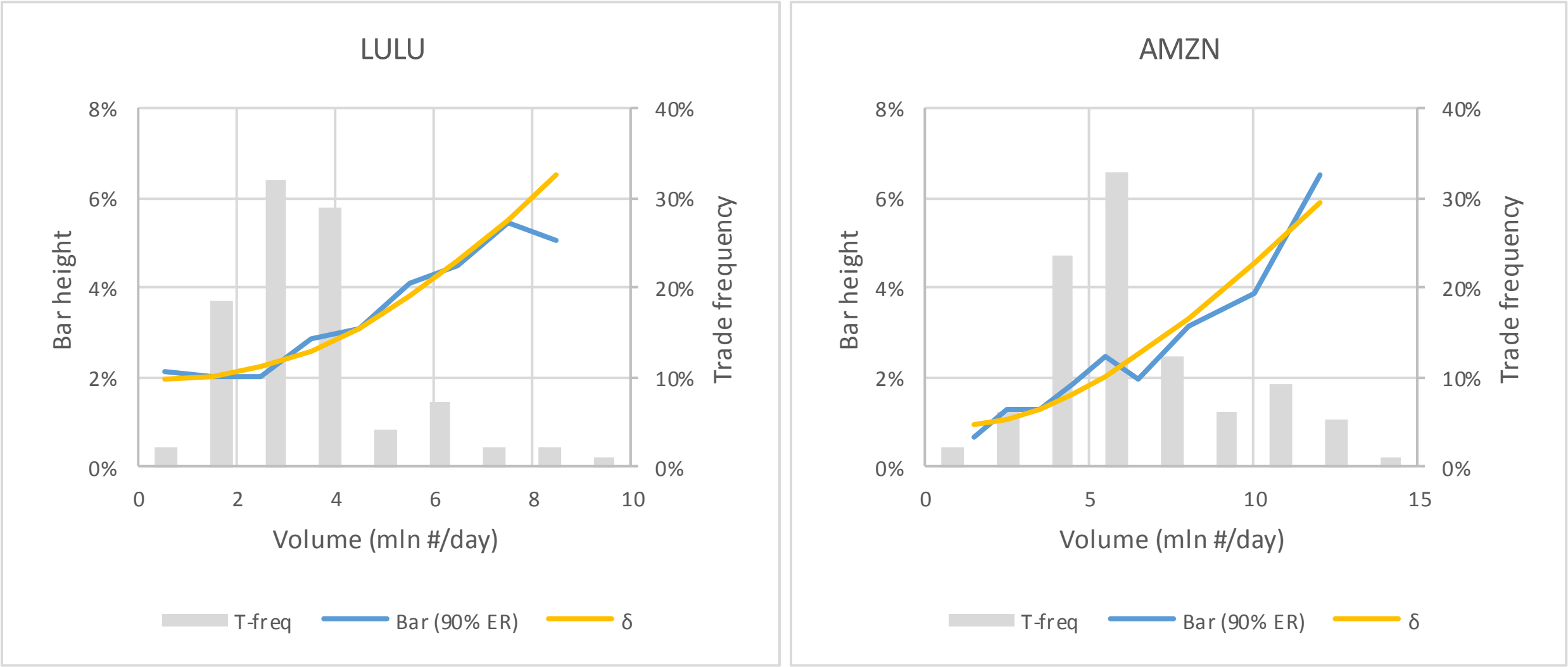


Fig. 13. Spread-volume curves based on daily high-low bars for LULU and AMZN. Lines correspond to execution rate (ER) 90%. Histogram represents the frequency of trading volume.

## 4. Spread control and market maker's profit optimization

Armed with functional dependence $\delta(v)$ we can now approach the problem of spread control and optimization. Market maker's P/L consists of two major components: spread revenue and inventory P/L. Spread revenue comes from executing buy and sell orders and keeping the price difference. Inventory P/L is the result of mark-to-market of the inventory held on market maker's book between buying and selling.

Since market makers essentially bet against price direction, that mark-to-market usually produces a loss. While inventory P/L is an extremely important component, that can substantially distort the net P/L profile, it has a substantially different nature from spread revenue and firms have various approaches dealing with it. Here we will focus on the spread revenue part.

If market maker executes bid and ask quotes with execution rate $r$ on a security that trades at volume $v$, then the market maker's turnover is $rv$. Execution costs and rebates $\alpha$ are usually proportional to turnover, so if quoted spread is $\delta$, earnings per round trip are $\delta - \alpha$. The spread P/L over the period is then equal

$$P/L = 0.5\, rv(\delta - \alpha) \tag{42}$$

Here $r$ is a function of $\delta$, and factor 0.5 reflects the fact that the security has to be bought and sold. An example of the resulting curve is shown in Fig. 14. The question is: what is the optimal operating spread that maximizes P/L? How should it be adjusted depending on current volume in order to guarantee maximum P/L? And are there conditions under which trading should be halted?

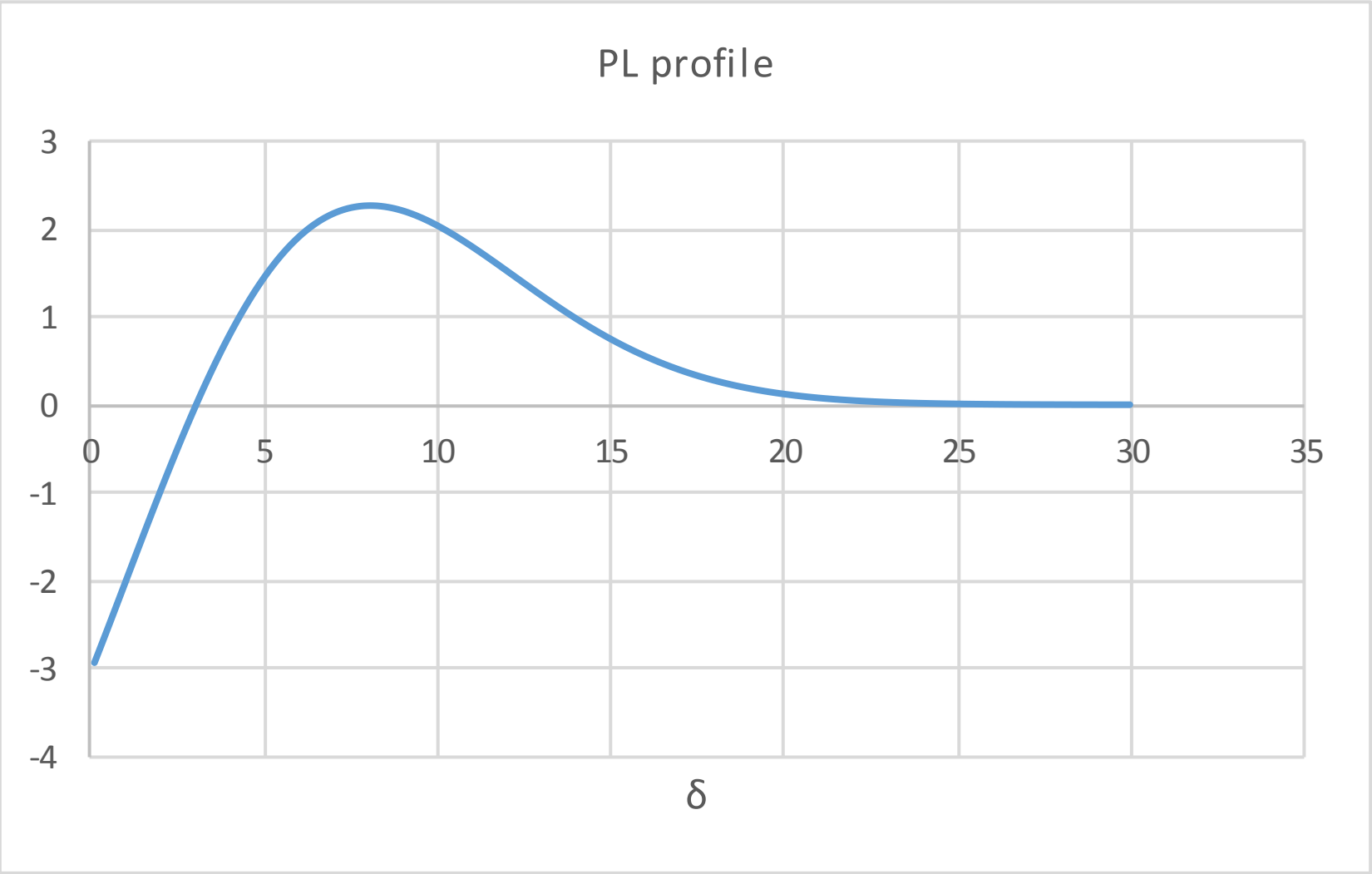


Fig. 14. PL dependency on operating spread. Very low $\delta$ is insufficient to cover trading expenses. As we widen it, profit grows reaching a maximum. With too large $\delta$ profit drops due to insufficient turnover.

In order to solve this problem, we need to establish relationship between $r$ and risk aversion level. This can be done approximately but quickly if we assume that $\lambda$ and $\rho$ scale similarly, so that $\rho \sim \lambda$. This way there is only one control parameter and $\delta \sim \lambda$. We know from [2] that probability distribution of spread can be approximated as $p(\delta) \sim \delta\, e^{-\left(\frac{\delta}{\delta_0}\right)^2}$. Since under our assumption $\delta \sim \lambda$, similar relation holds for $p(\lambda)$:

$p(\lambda) = 2\frac{\lambda}{\lambda_0^2}e^{-\left(\frac{\lambda}{\lambda_0}\right)^2}$, where $\lambda_0$ is a some constant. Execution rate is then the cumulative portion of the probability distribution that falls outside of range specified by $\lambda$:

$$r(\lambda) = \int_{\lambda}^{\propto} p(\lambda')d\lambda' = e^{-\left(\frac{\lambda}{\lambda_0}\right)^2} \quad (43)$$

Constant $\lambda_0$ can be calibrated to match the execution profile: $r(\lambda_0) = r_0$. Approximation Eq. (43) relies on a number of assumptions that need to be satisfied before applying it in practice. However, it can be easily modelled with the same framework for more complex situations. An example of execution rate with $\lambda_0 = 3$ is shown in Fig. 15.

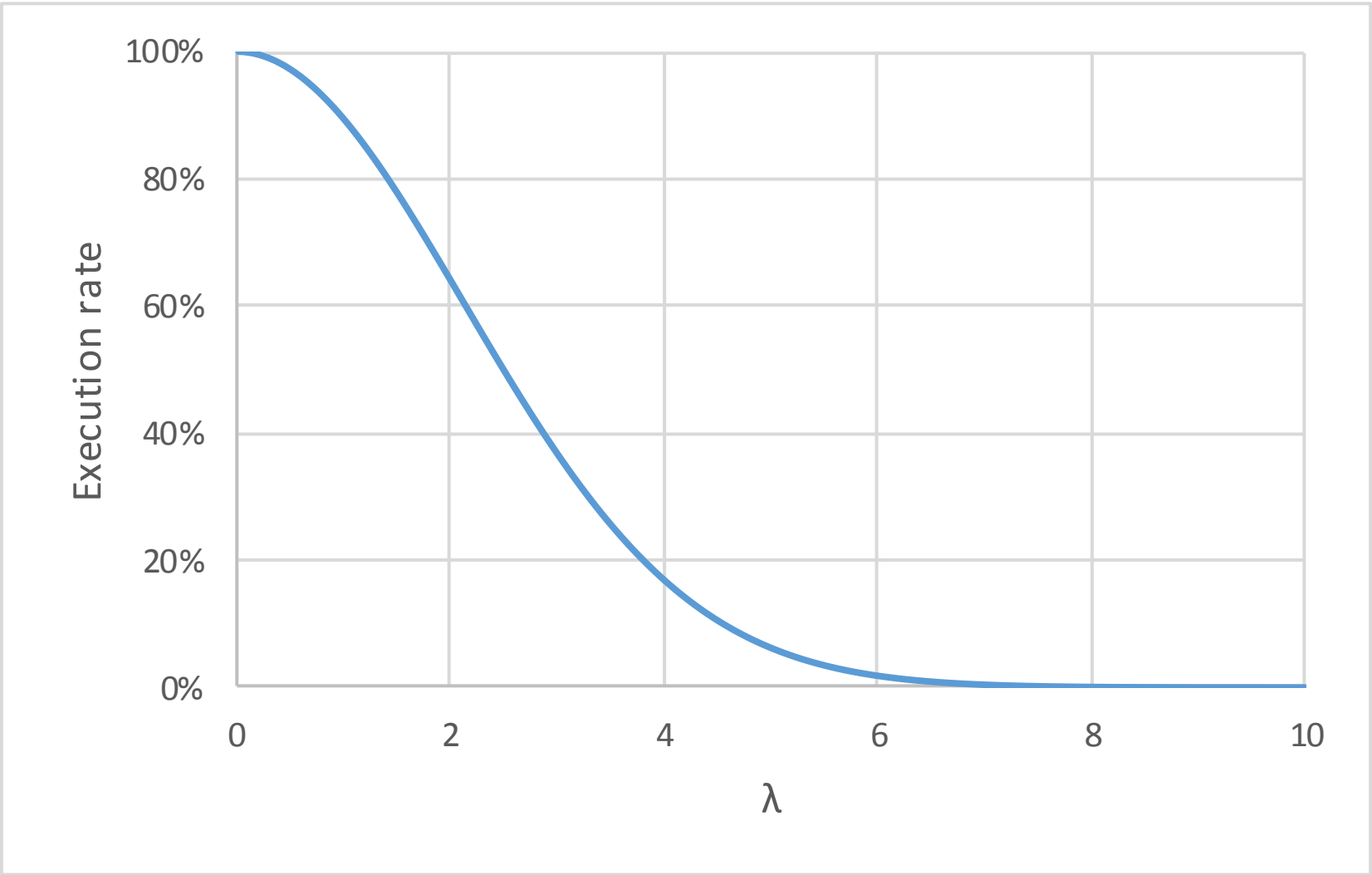


Fig. 15. Execution rate dependency on $\lambda$.

Optimal execution rate can then be found from solving the following equation with respect to $r$:

$$\delta - \alpha + r\frac{\partial\delta}{\partial r} = 0 \quad (44)$$

As an example, optimization results are shown in Figs. 16 and 17. We used $a = 10$ for spread, so its minimum value is 3, and commissions $\alpha = 3$, to better show the adjustments produced by the model when spread approaches commission. We see that in order to maximize profit the model requires operation at almost 2 times wider spread than the bid-ask spread, Figs. 16a and 17a, implying execution rate in the 45-55% range, Fig. 16b.

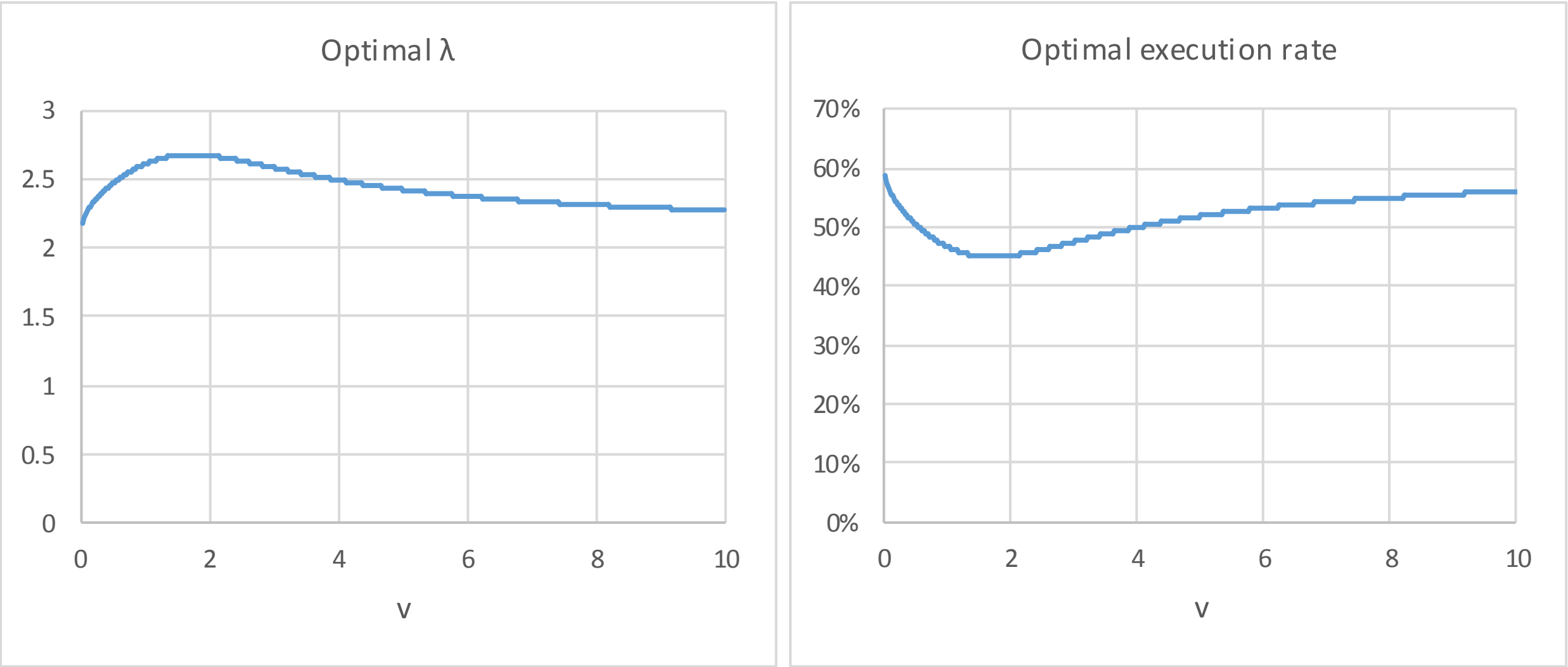

(a) (b)

Fig. 16. $\lambda$ and $r$ dependency on trading volume for bid-ask quoting.

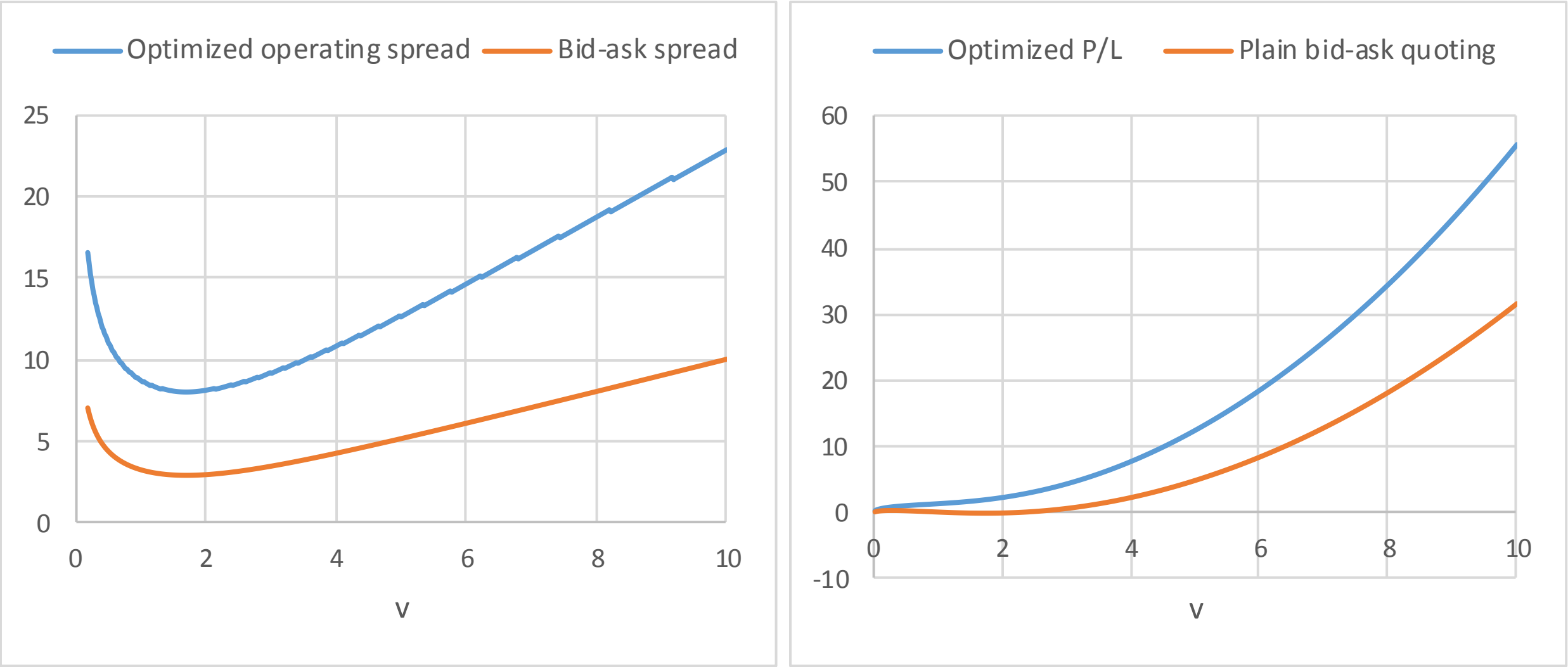

(a) (b)

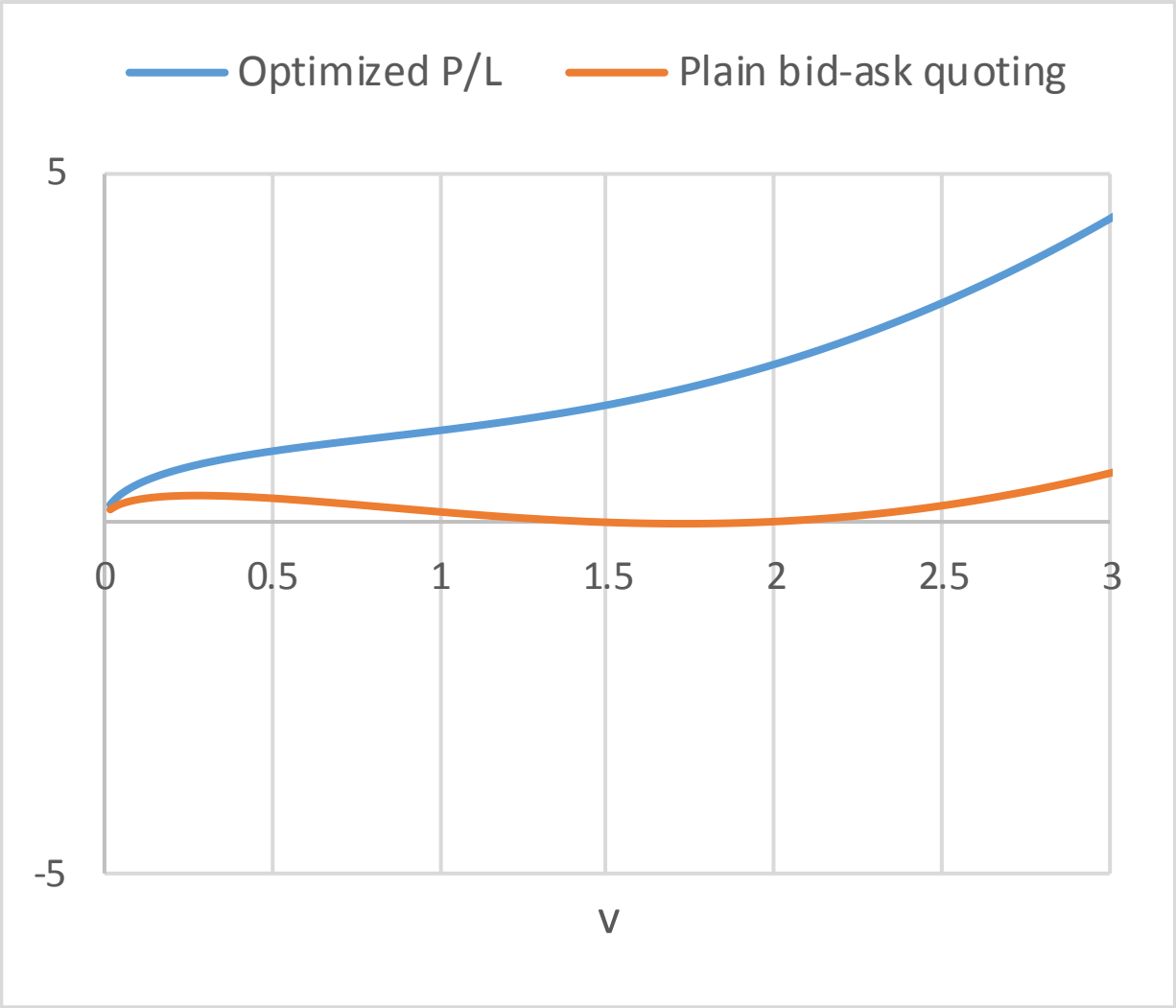

(c)

Fig. 17. (a) operating spread vs bid-ask spread, (b) comparison of P/L obtained with quoting at optimal operating spread vs. quoting at bid-ask spread, and (c) the same for the most tradable region, zoomed.

Behavior suggested by the model is opposite to that of buy-side. It proposes a more aggressive quoting (a) when trading volume is low and liquidity is limited, which is when market makers are most needed, and (b) when volume is large, which usually happens at times of large volatility, when the buy-side isn't sure about prices. Execution rate has to be lowered when bid-ask spread is around its minimum values, which helps to reduce the effect of the commissions. Optimized profit is shown in Figs. 17b and zoomed in Fig. 17c, versus the P/L made with plain bid-ask spread quoting. Similar charts are shown for bar-based quoting in Figs. 18, and 19.

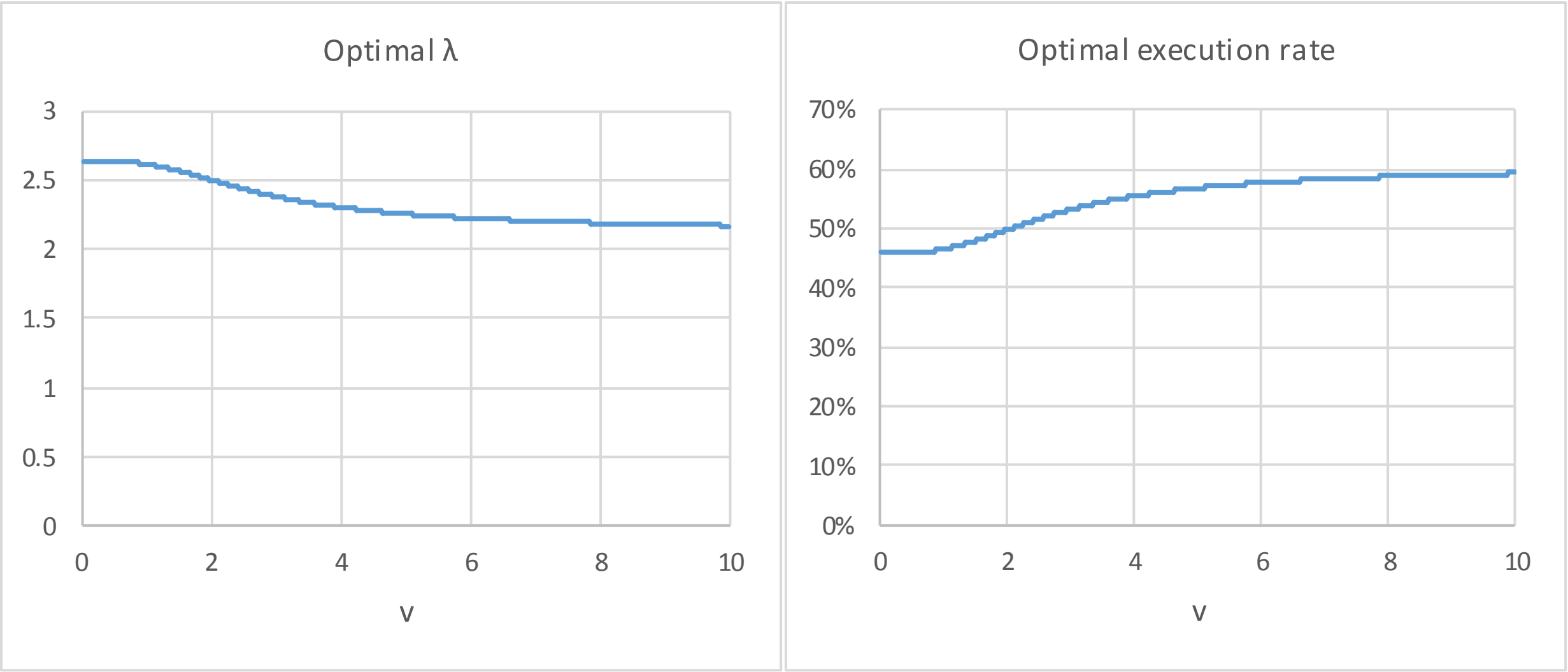

Fig. 18. $\lambda$ and $r$ dependency on trading volume for bar quoting.

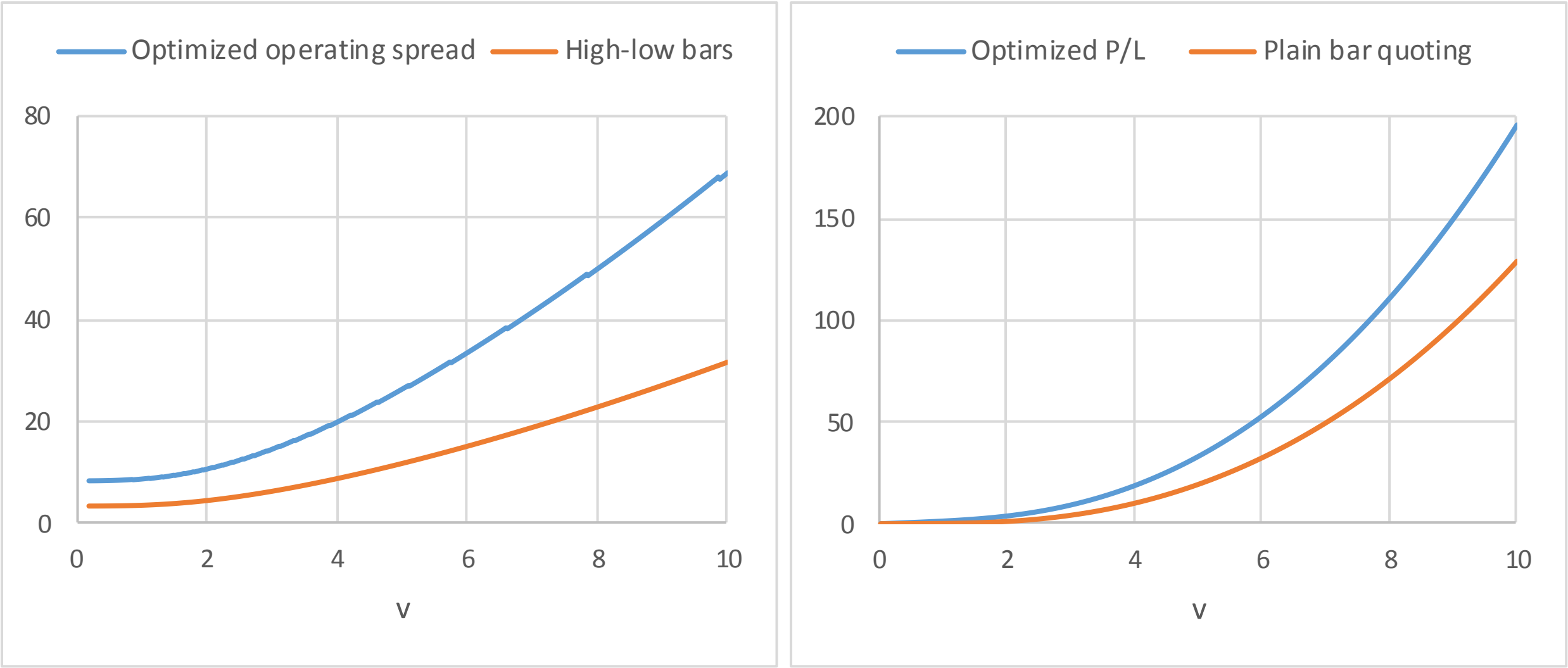

(a) (b)

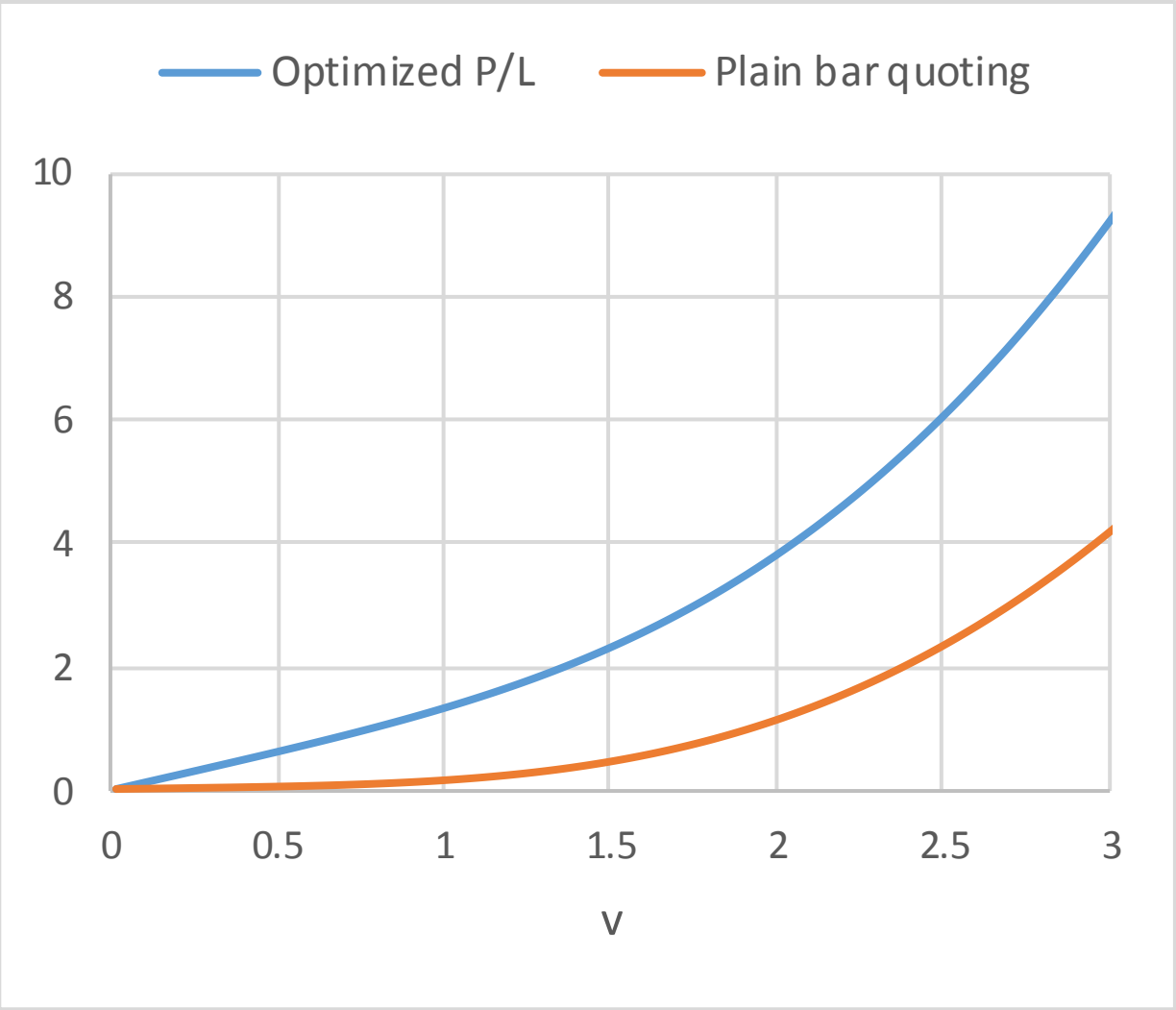

(c)

Fig. 19. (a) operating spread vs bars, (b) comparison of P/L obtained with quoting at optimal operating spread vs quoting at bars, and (c) the same for the most tradable region, zoomed.

With spread management provided by this framework, market makers can optimize their P/L and keep it optimized in changing market conditions. We see that large turnover does not necessarily mean maximum P/L, and that it may be beneficial to be less aggressive on execution. In commitments with minimum turnover requirements it provides a basis for rebate and commission negotiation.

Other versions of the considered problem are possible. For example, some desks might consider optimizing a utility function that maximizes the P/L, but penalizes for volatility or drawdown, limits inventory size, or sets any other restrictions that risk management department imposes on the desk.

## 5. Conclusions

In this work we established the relation of spread to microstructural parameters using basic stochastic considerations, and more generally, using the quantum coupled-wave model. We find that spread forms as a combination of two factors: the liquidity price, arising from price uncertainty due to liquidity limitations, and the impact price, arising from price impact caused by order flow. When liquidity is limited adding more liquidity helps improve price accuracy and reduce spread, but after some point additional liquidity begins to deteriorate price. As a demonstration of this the bid ask spread first decreases with volume and then starts increasing after reaching a minimum. High-low bars display a different behavior: they start with their minimum value and only keep growing.

The bar time-scaling law is more complex than the traditional $\sim\sqrt{T}$ behavior for volatility. At small time scale the impact price component keeps spread from reaching zero. The $\sim\sqrt{T}$ behavior restores for large $T$ when bar size becomes unimportant compared to volatility.

Combining the scaling results for volume and time, we were able to model bars as a function of volume and time horizon: $\delta = \delta(v, T)$. Such model allows to quickly adjust spread to current volume and switch between quoting time horizons.

This model's main limitation is the two-level assumption. Its results could break when order book is deep (many levels are filled with orders) and securities transfer involves more than two levels. In such case a multi-level model has to be applied. Such model was described in [1]. Additionally, because trading statistics is different for bid-ask spread and for bars, coupled-wave model should be carefully applied at small time scales at which bars are comparable in size with the spread.

All these results are consistent with market data on intraday and daily levels, so overall, we can say that "quantum coupled-wave model" produces viable results. It is important that no assumptions about the security, company fundamentals, or market structure were made in deriving these results.

Results about spread behavior were applied to solve the market maker's profit optimization problem. We showed how by setting spread at optimal value the spread revenue can be maximized. That value does not always correspond to quoting straight best bid and ask prices, and may require an execution rate that is substantially lower than 100%. Understanding spread behavior allows market makers to dynamically manage operating spread and keep profiting in any market conditions.

This model opens new capabilities for financial institutions that are involved in market-making and securities dealing activities. Using this framework firms and trading desks can price securities, particularly ones with limited liquidity, measure risk associated with spread, react quickly to changing market conditions, and optimize their income. All these capabilities are extremely important when a trading desk's risk/return profile substantially depends on spread.